\documentclass[a4paper,twocolumn,accepted=2025-02-03]{quantumarticle}
\pdfoutput=1
\usepackage[english]{babel}
\usepackage[T1]{fontenc}
\usepackage[utf8]{inputenc}
\usepackage{amsmath, bbm}
\usepackage{amssymb}
\usepackage{physics}
\usepackage{graphicx}
\usepackage{xcolor} % coloring text
\usepackage[normalem]{ulem} % striking text under review
\usepackage[T1]{fontenc}
\usepackage{url}
\usepackage{hyperref}
\usepackage[numbers,sort&compress]{natbib}
\usepackage{float}

\newcommand\work{work}
\newcommand\refeq[1]{Eq.~(\ref{#1})}
\newcommand\reffig[1]{Fig.~\ref{#1}}

\makeatletter

\pdfpageheight\paperheight
\pdfpagewidth\paperwidth

\makeatother

\begin{document}
\title{Multipartite entanglement distribution in a topological photonic network}
\author{Juan Zurita}
\affiliation{Instituto de Ciencia de Materiales de Madrid (CSIC), Cantoblanco,
E-28049 Madrid, Spain}
\affiliation{Departamento de F\'isica de Materiales, Universidad Complutense de Madrid,
E-28040 Madrid, Spain}
\email{juzurita@ucm.es}

\author{Andr\'es Agust\'i Casado}
\affiliation{Instituto de Ciencia de Materiales de Madrid (CSIC), Cantoblanco,
E-28049 Madrid, Spain}
\author{Charles E. Creffield}
\affiliation{Departamento de F\'isica de Materiales, Universidad Complutense de Madrid,
E-28040 Madrid, Spain}
\author{Gloria Platero}
\affiliation{Instituto de Ciencia de Materiales de Madrid (CSIC), Cantoblanco,
E-28049 Madrid, Spain}
\begin{abstract}
In the ongoing effort towards a scalable quantum computer, multiple technologies have been proposed. Some of them exploit topological materials to process quantum information. In this work, we propose a lattice of photonic cavities with alternating hoppings to create a modified multidomain SSH chain, that is, a sequence of topological insulators made from chains of dimers. A qubit is then coupled to each boundary. We show this system is well suited for quantum information processing because topological transfer of photons through this one-dimensional lattice can entangle any set of qubits on demand, providing a scalable quantum platform. We verify this claim evaluating entanglement measures and witnesses, proving that bipartite and multipartite entanglement is produced, even in the presence of some disorder.
\end{abstract}
\maketitle
\global\long\def\a{\alpha}%
\global\long\def\b{\beta}%

\global\long\def\vc{\vartheta}%
\global\long\def\d{\delta}%
\global\long\def\e{\epsilon}%
\global\long\def\ve{\varepsilon}%
\global\long\def\f{\phi}%
\global\long\def\vf{\varphi}%
\global\long\def\g{\gamma}%
\global\long\def\i{\iota}%
\global\long\def\j{\chi}%
\global\long\def\k{\kappa}%
\global\long\def\l{\lambda}%
\global\long\def\m{\mu}%
\global\long\def\n{\nu}%
\global\long\def\p{\pi}%
\global\long\def\q{\psi}%
\global\long\def\r{\rho}%
\global\long\def\s{\sigma}%
\global\long\def\vs{\varsigma}%
\global\long\def\t{\tau}%
\global\long\def\u{\upsilon}%
\global\long\def\w{\omega}%
\global\long\def\x{\xi}%
\global\long\def\y{\eta}%
\global\long\def\z{\zeta}%
\global\long\def\C{\Theta}%
\global\long\def\D{\Delta}%
\global\long\def\F{\Phi}%
\global\long\def\G{\Gamma}%
\global\long\def\L{\Lambda}%
\global\long\def\P{\Pi}%
\global\long\def\Q{\Psi}%
\global\long\def\S{\Sigma}%
\global\long\def\U{\Upsilon}%
\global\long\def\W{\text{\ensuremath{\Omega}}}%
\global\long\def\X{\Xi}%
\global\long\def\bbZ{\mathbb{Z}}%
\global\long\def\uno{\mathbbm{1}}%
\global\long\def\sp{\boldsymbol{s^{\prime}}}%
\global\long\def\kk{|0\rangle_{\mathcal{L}}|0\rangle_{\g}|0\rangle_{\mathcal{R}}}%
\global\long\def\kt{\rangle}%
\global\long\def\sA{\mathcal{A}}%
\global\long\def\sL{\mathcal{L}}%
\global\long\def\sR{\mathcal{R}}%
\global\long\def\sS{\mathcal{S}}%
\global\long\def\sP{\mathcal{P}}%
\global\long\def\sH{\mathcal{H}}%
\global\long\def\sW{\mathcal{W}}%

\global\long\def\kk#1#2#3{ |#1\rangle_{\mathcal{L}}|#2\rangle_{\gamma}|#3\rangle_{\mathcal{R}} }%
\global\long\def\kkk#1#2#3{ |#2\rangle_{#1}|#3\rangle_{\gamma} }%
\global\long\def\vac{\varnothing}%
\global\long\def\cnot{\textrm{CNOT}}%

\global\long\def\down{\bigg\downarrow}%

\section{Introduction}
Quantum mechanics allows for the creation of a computer able to solve problems with a higher performance than classical ones \cite{Feynman1982, shor1994, grover1996}. Plenty of research nowadays deals with this statement, from the study of algorithms and metrics that prove quantum advantage \cite{boixo_isakov2018, Bravyi2018}, to the design of hardware primitives able to create said computer \cite{cirac1995,zhong_wang2020}. In this \work{} we focus on the latter, the physical elements of a quantum computer, and analyse them with a metric that is a necessary condition to quantum advantage: multipartite entanglement production \cite{nielsen2006}. A list of proposals for quantum computers includes superconducting circuits \cite{blais_grimsmo2021}, ion or cold atoms traps \cite{pogorelov_feldker2021}, photonic systems,  quantum dots, and topological materials, to name but a few. It is unclear which platform will have the most impact, but the history of classical computing teaches us that multiple implementations coexisted and all are worth researching. Here we focus on a topological photonic lattice coupled to color center emitters, given the outstanding progress taking place on those respective areas of research, that we briefly introduce now.

\begin{figure}[!h]
\includegraphics[width=1\columnwidth]{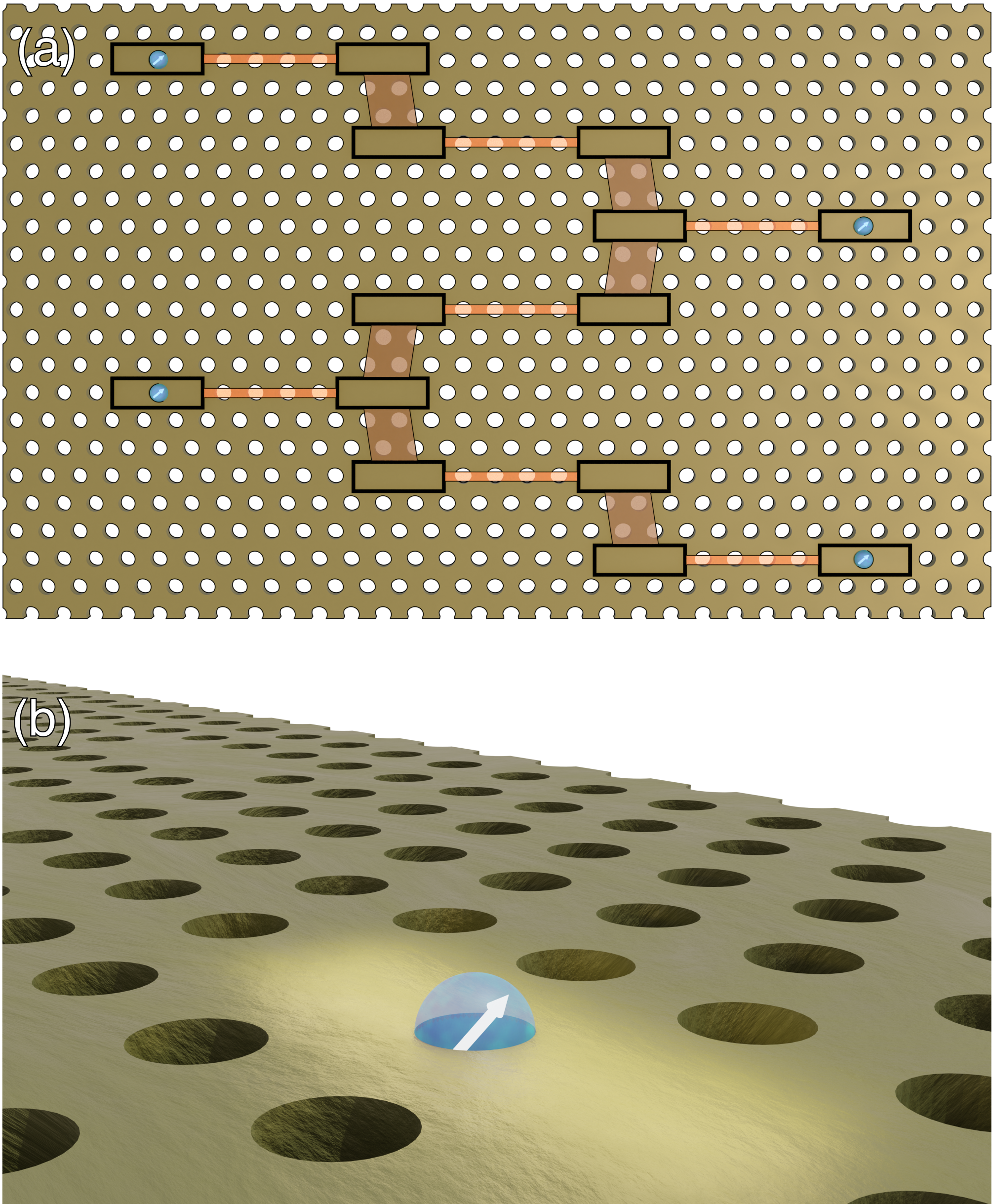}\caption{\label{fig:PhotoCavities}(a) A three-domain, 14-cavity Si photonic lattice. The four color center qudits are represented as blue spheres. (b) Close-up of a color center qubit in a cavity (highlighted).}
\end{figure}

Topological phases of matter are an interesting quantum platform because of their robustness against some kinds of disorder. This protection helps with the survival of quantum properties for longer times, the main limiting factor in quantum computing applications. One of the applications topological matter finds in quantum technologies is the use of 1D symmetry-protected edge states to transfer information across the material \cite{Bello2016,Bello2017,Lang2017,Longhi2019,Longhi2019a,Yuce2019,Perez-Gonzalez2019,DAngelis2020}. This transfer can take place along several regions with different topological characteristics, that is, along multiple domains. Recently, it was shown that a multidomain approach drastically accelerates transfer \cite{Zurita2023}, which greatly increases the technological potential of this phenomenon. In that regard, photonic lattices are experiencing an outstanding growth in the number and complexity of topological models they can implement \cite{Malkova2009,Wang2009,Hafezi2013,Weimann2017,St-Jean2017,Zhao2018,Yang2018,Huda2020,Zurita2020,Vega2021,Gong2021,Pernet2022,Rosiek2022,Kim2022a} and use in applications like lasing \cite{Bandres2018,Contractor2022}, sensing \cite{Parto2023} or, as mentioned, quantum information processing \cite{Blanco-Redondo2018,Hu2020,Tschernig2021, Dai2022}. In particular, multidomain SSH-like models have been implemented experimentally in several platforms  \cite{Blanco-Redondo2016,Zhao2018,Huda2020,Pernet2022}.

Another reason to turn our attention to photonics is their ability to couple to emitters, forming highly scalable platforms that are created using advanced fabrication techniques \cite{Khoury2022}. Recent breakthroughs have been achieved like the production of high-brightness G-centers and the control of their position \cite{Hollenbach2020,Hollenbach2022}, the coupling of G-centers to a photonic cavity in a single-photon basis \cite{Baron2022,Redjem2023}, and the creation of vacancies or W-, T- and G-centers coupled to waveguides \cite{Tait2020,MacQuarrie2021,DeAbreu2022,Prabhu2023,Lukin2023}.

The advances cited so far have led to theoretical works about whether these emitters can be used as qubits of a quantum computer, coupled to each other through the edge states of a photonic topological system \cite{Lang2017,Yan2021,Hollenbach2020,Vega2021,Guimbao2021}. In this \work{} we address this question, that is, we study whether coherent operation of multi-qubit states is possible. Furthermore, we propose protocols for coherent operation that scale well with the number of qubits. To show this we propose a nearly one-dimensional system, and prove that it produces entanglement between any pair or bigger subset of qubits in experimentally feasible settings, proving that coherent processes mediated by the edge states overcome dissipation and decoherence.

The structure of this \work{} is as follows: in Section \ref{sec:system}, we introduce the system, consisting of a quasi-1D topological photonic lattice, with qubits placed at its interfaces. We also propose an experimental implementation for it, using color center qubits in a silicon cavity lattice. Next, we demonstrate in Section \ref{sec:Bell} how the topological transfer of a photon through the lattice can be used to prepare a Bell state between arbitrary pairs of these qubits. To quantify the performance of the protocol we study the concurrence between the target qubits in the presence of different kinds of disorder. We then use similar topological transfers in Section \ref{sec:Multi} to prepare the W and GHZ states between three qubits and evaluate the performance of the protocol using genuine multipartite entanglement witnesses in the same parameter regimes as with Bell states. In Section \ref{sec:Loss}, we explore the main sources of decoherence and loss affecting our protocols.  Finally, we state our conclusions about the usability and scalability of this system as a quantum computing platform, together with possible future lines of research.

\section{Description of the system}\label{sec:system}
We consider a cavity lattice made up of several topological regions, with a number of qubits coupled to the boundaries between them. The full system will be described by:
\begin{equation}
    \sH_T = \sH_\g + \sH_Q + \sH_{Q\g},
\end{equation}
where $\sH_\g$ describes the topological photonic lattice, $\sH_Q$ describes the qubits and $\sH_{Q\g}$ governs the interactions between them.

\subsection{Topological photonic lattice}
We first focus on the topological lattice. The simplest topological insulator, the SSH chain, consists of a chain of atoms with alternating tunnelling amplitudes $v$ and $w$. Its end states are protected against disorder in said amplitudes as long as the intracell links $v$ are weaker than the intercell ones $w$. We consider a lattice of photonic cavities made up of several SSH-like domains, where each new domain is coupled to the penultimate site in the previous one, creating domain walls with additional sites (or \textit{stubs}) coupled to them [see Fig. \ref{fig:PhotoCavities} (a)]. We will refer to this lattice as the stub-SSH (SSSH) model. As in the SSH chain, each unit cell has two sites, $A$ and $B$, that we label with $\alpha$. They take on alternating roles in each domain. The Hamiltonian describing the model is:
\begin{align}
    \sH_\g = -u_0 (b_1^{\dagger} a_1 + a_1^{\dagger} b_1) - \sum_{k=1}^N \sH_\g^{(k)}, \label{eq:Hph}
\end{align}

where $\sH_\g^{(k)}$ is the Hamiltonian belonging to the $k$-th domain. It takes a different form for odd and even values of $k$. For odd $k$, it is given by:

\begin{align}
    &\sH_\g^{(k)}=
        \sum_{j=1}^{\ell/2}w a_{j_{0}(k)+j}^{\dagger}b_{j_{0}(k)+j-1}+ \\
 \sum_{j=1}^{\ell/2-1}\! v_{k}&b_{j_{0}(k)+j}^{\dagger}a_{j_{0}(k)+j} 
 +u_{k}b_{j_{0}(k)+\ell/2}^{\dagger}a_{j_{0}(k)+\ell/2} + h.c.,\nonumber
\end{align}
while for even $k$, the $a$ and $b$ operators are swapped:
\begin{align}
    &\sH_\g^{(k)}=
        \sum_{j=1}^{\ell/2}w b_{j_{0}(k)+j}^{\dagger}a_{j_{0}(k)+j-1}+ \\
 \sum_{j=1}^{\ell/2-1}\! v_{k}&a_{j_{0}(k)+j}^{\dagger}b_{j_{0}(k)+j} 
 +u_{k}a_{j_{0}(k)+\ell/2}^{\dagger}b_{j_{0}(k)+\ell/2} + h.c.\nonumber
\end{align}

In the expressions above, $j_0(k)=k\ell/2+1$ is the unit cell of the $k$-th domain wall and operator $a_{j}^{\dagger}$ $(b_{j}^{\dagger})$ creates a photon in the cavity in unit cell $j$ and sublattice $\alpha=A(B)$. The $v_k$ and $w$ hopping amplitudes form the main SSH domains, while the $u_k$ amplitudes connect the cavities at the ends and stubs to the rest. The lattice is determined by its number of SSH-like domains $N$ and their domain length $\ell$. An additional cavity is coupled to each domain wall, such that each domain $k=1,\ldots,N$ has $\ell$ inner sites and is bounded by two extremal sites. We consider even values of $\ell$. The total number of cavities is $N_{cav}=N\ell+2$.
%CAUTION: $\ell_{programs}=\ell_{here}-1$. I changed it here because it makes all formulas easier. Here it's always even, in the programs it's always odd.

\begin{figure}[H]
\centering
\includegraphics[width=0.7\columnwidth, height=0.5\textheight, keepaspectratio]{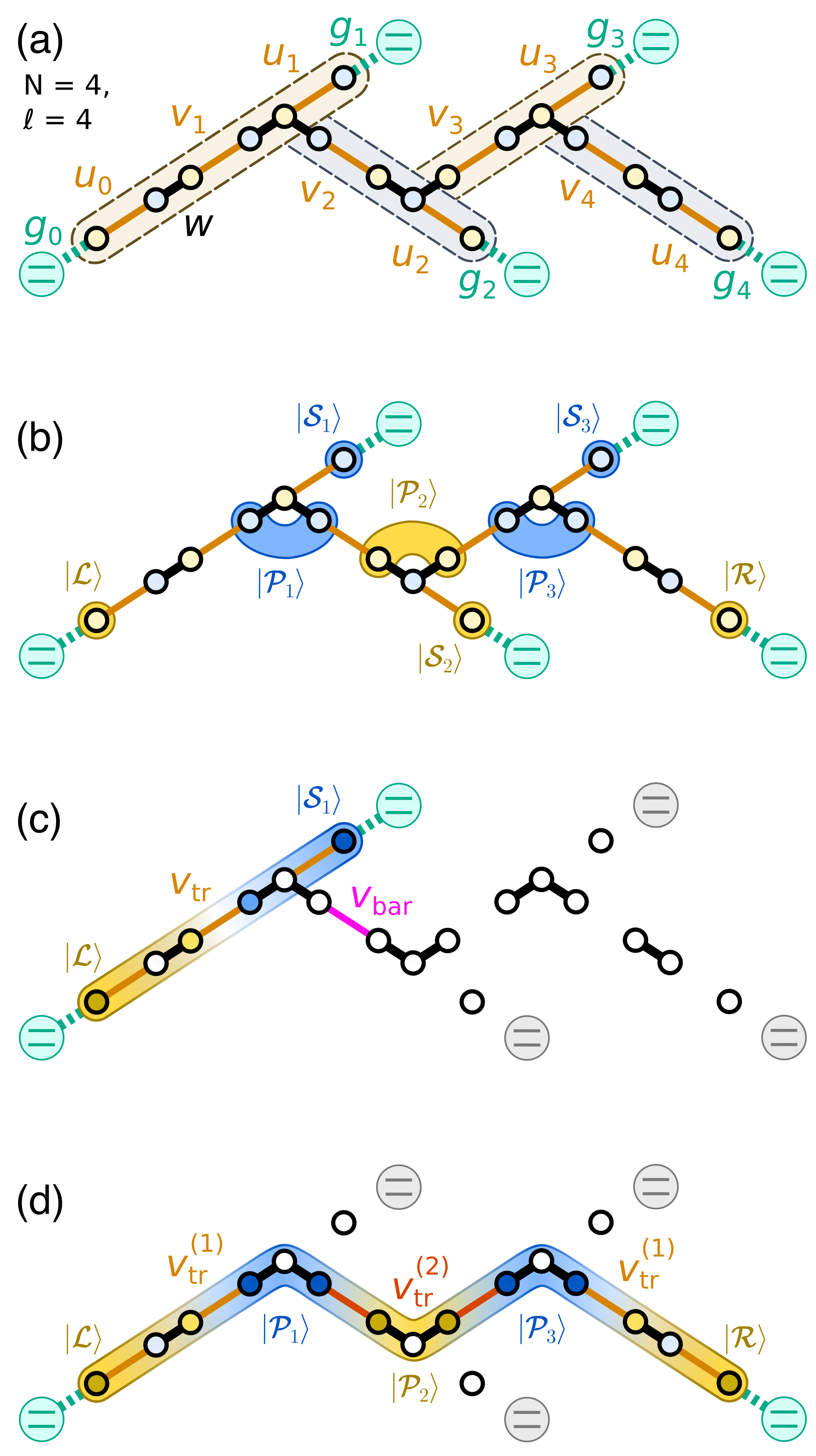}\caption{\label{fig:model}(a) Photonic cavity lattice with $N=4$ domains and $\ell=4$ cavities per domain. Qubits (green) are coupled to the ends and domain walls, with an amplitude of $g_{k}$. The links used as control parameters are shown in orange: we use $u$ for extremal links, $v$ for the rest. (b) Topologically protected states in this system. Two chiral states, $\mathcal{S}$ and $\mathcal{P}$, exist at each wall. States in odd (even) boundaries are confined to sublattice $B$ ($A$), and shown in blue (yellow). (c) Transfer between the left end and the first wall. The parameter $v_{\mathrm{tr}}<w$ allows the transfer, while $v_{\mathrm{bar}}\gg w$ stops it. (d) End-to-end transfer. The $\mathcal{P}$ states are involved now. Two different pulses are used for the extremal ($v_{\mathrm{tr}}^{(1)}$) and central ($v_{\mathrm{tr}}^{(2)}$) domains to achieve a high-fidelity transfer.}
\end{figure}

\subsection{Coupled qubits}
Qubits can be coupled to any number of extremal cavities, as shown in Fig. \ref{fig:PhotoCavities} (a,b), and we will label them starting at 0. They will act like photon emitters with the frequency of the cavity. In one of the protocols, we will extend the qubits to qutrits, so we keep the notation general. The coupling between the qudits and the cavities housing them is, in the rotating wave approximation:

\begin{equation}
\mathcal{H}_{Q\g}=\sum_{k=0}^{p-1}g_{k}c_{j(k),\alpha(k)}|s_{u}\kt\langle s_{l}|+h.c.,
\end{equation}
where $j(k),\alpha(k)$ labels the cavity where the $k$-th qudit is located, $c_{j(k),\alpha(k)}$ destroys a photon in that cavity, $p$ is the number of qudits, $g_k$ is the coupling coefficient for the $k$-th qudit, and
$|s_{l}\kt$ and $|s_{u}\kt$ are the lower and upper qudit states
connected by the cavity-mediated transition, with $s_{l,u}=0,\ldots,d-1$, where $d$ is the dimension of the qudits used.

The odd domains in this model are SSH chains, while the even ones can be obtained as a limiting case of the extended SSH model studied in \cite{Perez-Gonzalez2019}. The use of domains with topological invariants of $\pm 1$ (see Appendix \ref{sec:TopoProt}) provides domain walls with two states each, one of them can be used to interface with a qubit while the other is used to transfer information along the wall. This allows us to create a 1D network of qubits with full connectivity. A system with $N$ domains then has $2N$ boundary states, that are topologically protected against off-diagonal disorder as long as $w$ is greater than all $v$-type (intra-domain) and $u$-type (stub) links, as shown in Fig. \ref{fig:model} (a,b). Some of these states are mainly localized in a single cavity and can thus be easily coupled to a qubit (the left and right states $|\sL\kt$, $|\sR\kt$ and the $|\sS_k\kt$ states inside the domain walls), while others ($|\sP_k\kt$) can be used to transfer information along a wall. These last states disappear if we set $v_{k}>w$ in one of their contiguous domains. We show this, as well as the extent of the topological protection, in Appendix \ref{sec:TopoProt}.

All edge states that are not decoupled from the rest of the system (e.g. by setting their $u_k=0$) will hybridize, forming bonding and antibonding eigenstates. The topological protection ensures that they will stay decoupled from all bulk states. The effective overlap between each pair of states depends on the links between them, given that these dictate their decay length.

For this reason, a transfer protocol between the two ends of all the hybridized states can be designed by tuning some of the hopping amplitudes. In fact, by tuning only the weaker hopping amplitudes, marked in orange in Fig. \ref{fig:model} (a), the dynamics in this subspace can be fully controlled. Using the right pulses, a photon initially on the leftmost state of the set of hybridized states can always be made to end up on the rightmost one, as explored in \cite{Zurita2023}. The effective Hamiltonian of the protected subspace in the SSSH chain is analogous to that of the Creutz ladder in the mentioned work. We show the pulses we used in Appendix \ref{sec:pulses}. This transfer process, which we will refer to as $T$, is always topologically protected and can be established between any two domain boundaries.

These pulses are also used to completely localize the initial and final states of the transfer in a single cavity, in order for the qubit-end mode coupling to be perfect. If no parameters could be tuned at all, end-to-end transfers would still be possible, although their fidelity would be smaller because the end modes would not coincide exactly with the end cavity states.

Apart from connectivity, there is an additional advantage to using domains. As discussed in the literature \cite{Lang2017,Zurita2023}, a downside of bidirectional transfer protocols in 1D topological insulators is the exponential scaling of their duration as a function of distance, which in turn allows small symmetry-breaking perturbations to become catastrophic for the performance of the protocol. This is caused by their exponential decay into the bulk, which determines their effective overlap. This problem can be solved by adding intermediate domain walls between the ends of the model, as shown in \cite{Zurita2023}. This makes the transfer duration grow only linearly with distance.

\subsection{Experimental proposal} \label{ssec:exp} The cavity lattice can be built using a triangular lattice of holes in a slab of silicon, where the absence of holes creates a cavity, as shown in Fig. \ref{fig:PhotoCavities} (a). The hopping amplitude between contiguous cavities can be modified by changing the medium between cavities, by driving the cavity frequencies \cite{Liao2010} or using ancillary cavities (either containing impurities \cite{Qin2016} or with high frequencies \cite{Liao2010,Mukherjee2018}), or even by implementing nanowire-induced movable cavities, as shown experimentally in \cite{Birowosuto2014b}. As we explain below, the ability to tune these amplitudes is not necessary to implement the two-qubit protocol, although it does increase the resulting fidelity.

The qubits can be impurities in the silicon like G- or T-centers, located within the appropriate cavities, as shown in Fig. \ref{fig:PhotoCavities} (b). The qudit-cavity coupling parameters, $g_{k}$, could be modified by tuning the energies of the impurity with an electrical gate in or out of resonance with the cavity \cite{Hollenbach2020}, by tuning the cavity frequency as mentioned above, or otherwise by using an indirect transition with an auxiliary laser \cite{Lang2017}.

In Section \ref{sec:Loss}, we discuss the main challenges our protocols would face in this platform, and how they could be overcome or mitigated using the topological features of the lattice.

\section{Bell state preparation}\label{sec:Bell}
\subsection{Preparation protocol}
In this section, we use this photonic multidomain topological system to prepare maximally entangled states on two qubits, i.e. Bell states. In order to explain the protocol, we need to define some notation. Let $n_d$ be the total number of domains and $\ell$ the number of sites per domain. As explained above, qubits sit on the stubs between domains. Let $R$ and $L$ stand for the rightmost and leftmost qubits, which are the furthest apart and so the hardest to entangle. Therefore, we describe the protocol for these qubits only, given that the protocol for any other pair is analogous. The ground and excited states of qubit $L$ are $\ket{0}_L$ and $\ket{1}_L$, respectively, and we will refer to its coupling to the photonic cavity stub here as $g_L$.
Furthermore, since coherent single-photon emission has been reported experimentally for color-center qubits \cite{Baron2022,Redjem2023}, we consider $\pi$- and $\pi/2$-pulses to be an accessible operation between qubit $L$ and its cavity\footnote{We use sine-shaped pulses for all $g$ parameters, see Appendix \ref{sec:pulses} for details.}. We label the latter unitary as $(\Pi/2)_{01}^{(L)}$. We use similar notation for qubit $R$. The vacuum state of the photonic lattice is $\ket{\varnothing}_\g$. The topological transfer of a photonic excitation from the leftmost cavity to the rightmost through the edge states is $T_{LR}$.

With this notation now fixed, we describe the protocol that prepares a Bell state from a separable state as shown in Fig.~\ref{fig:2Q}~(a). The steps consist of: (i) Prepare the initial state $\ket{1}_L\otimes\ket{0}_{R} \otimes \ket{\varnothing}_\g$. (ii) Execute a $(\Pi/2)_{01}^{(L)}$ so that the excited qubit emits half its probability amplitude to the photonic lattice. Note it does so with a $-i$ relative phase. (iii) Perform a full transfer $T_{LR}$ so the photonic amplitude is now at the rightmost cavity. Note that the relative phase changes again. (iv) Execute a $\pi$-pulse on $g_{R}$, that is, $\Pi_{01}^{(R)}$. The final qubit state will be 
\begin{align}
    \ket{\Phi}_{LR} = \frac{1}{\sqrt{2}}\left[
        \ket{10} + e^{i\theta} \ket{01}
    \right],
    \label{ideal-bell-state}
\end{align}
where the relative phase is 
\begin{equation}
e^{i\theta} = [(-1)^{\ell/2+1}i]^{n_{d}}.
\end{equation}
See Appendix \ref{sec:phases} for a detailed derivation of this relative phase\footnote{This formula is related to those given in \cite{Zurita2023} for the SSH chain, where $\ell$ was defined differently and $n_{d}=n_{w}+1$.}, and Appendix \ref{sec:steps} for the intermediate steps in this and all protocols.

\subsection{Performance against disorder: concurrence}
While the calculations reported so far demonstrate that this system generates entanglement in the ideal case, we need to explore the impact of disorder if we are to consider the system a scalable quantum platform. First of all, the final state will no longer be \refeq{ideal-bell-state}. Moreover, the final state has a nonzero probability amplitude of having a leftover photon in the photonic lattice, leading to decoherence in the state of the qubits. In this case, entanglement certification is not as simple as inspecting the state in \refeq{ideal-bell-state}, so instead we turn to concurrence, an entanglement measure \cite{Hill1997}, to perform this task. We remind the reader that, if a two-qubit state has high concurrence, any latter state with lower concurrence may be prepared from the former with only local operations and classical communication (LOCC); in particular, without using the photonic lattice. In this sense, Bell states have maximal concurrence. We now turn our attention to the disorder model used, and then show that the protocol can produce high-concurrence states despite disorder.

We consider quasistatic disorder, that is, constant in time for each simulation, on each matrix element of the tight-binding Hamiltonian:
\begin{equation}
\mathcal{H}_{j,\alpha;j',\alpha'}=\mathcal{H}_{j,\alpha;j^{\prime},\alpha^{\prime}}^{(0)}+h_{j,\alpha;j^{\prime},\alpha^{\prime}}^{(\sigma)},
\end{equation}
where each $h_{j,\alpha;j^{\prime},\alpha^{\prime}}^{(\sigma)}$ is an independent random variable sampled once per realization with a normal probability distribution centered at zero and with a standard deviation of $\sigma$, which corresponds to the intensity of disorder. We tune the diagonal and off-diagonal levels of disorder independently to showcase the topological protection of our system towards off-diagonal (OD) symmetry-preserving disorder, as well as to check the impact of general (G) symmetry-breaking noise.

There is a second feature of the system with regard to noise protection. In \cite{Zurita2023} it was shown that the use of topological domains drastically accelerates transfer protocols---thus reducing the effects of disorder---and increases connectivity. To clarify this quantitatively, we consider two different kinds of simulations: a single, long SSH chain between the qubits with 14 cavities in a line, and the four-domain lattice shown in Fig. \ref{fig:model} (a), with 15 cavities between the qubits. The transfer time goes from $t_{\mathrm{tr}}=304.0/w$ for one domain (with a preparation time of $t_{\mathrm{prep}}=25/w$) to $t_{\mathrm{tr}}=45.3/w$ for four domains (with $t_{\mathrm{prep}}=20/w$). This speed-up, which is exponential in the length of the chain, is crucial in reducing the effects of diagonal disorder on entanglement production.

\begin{figure}
\includegraphics[width=1\columnwidth]{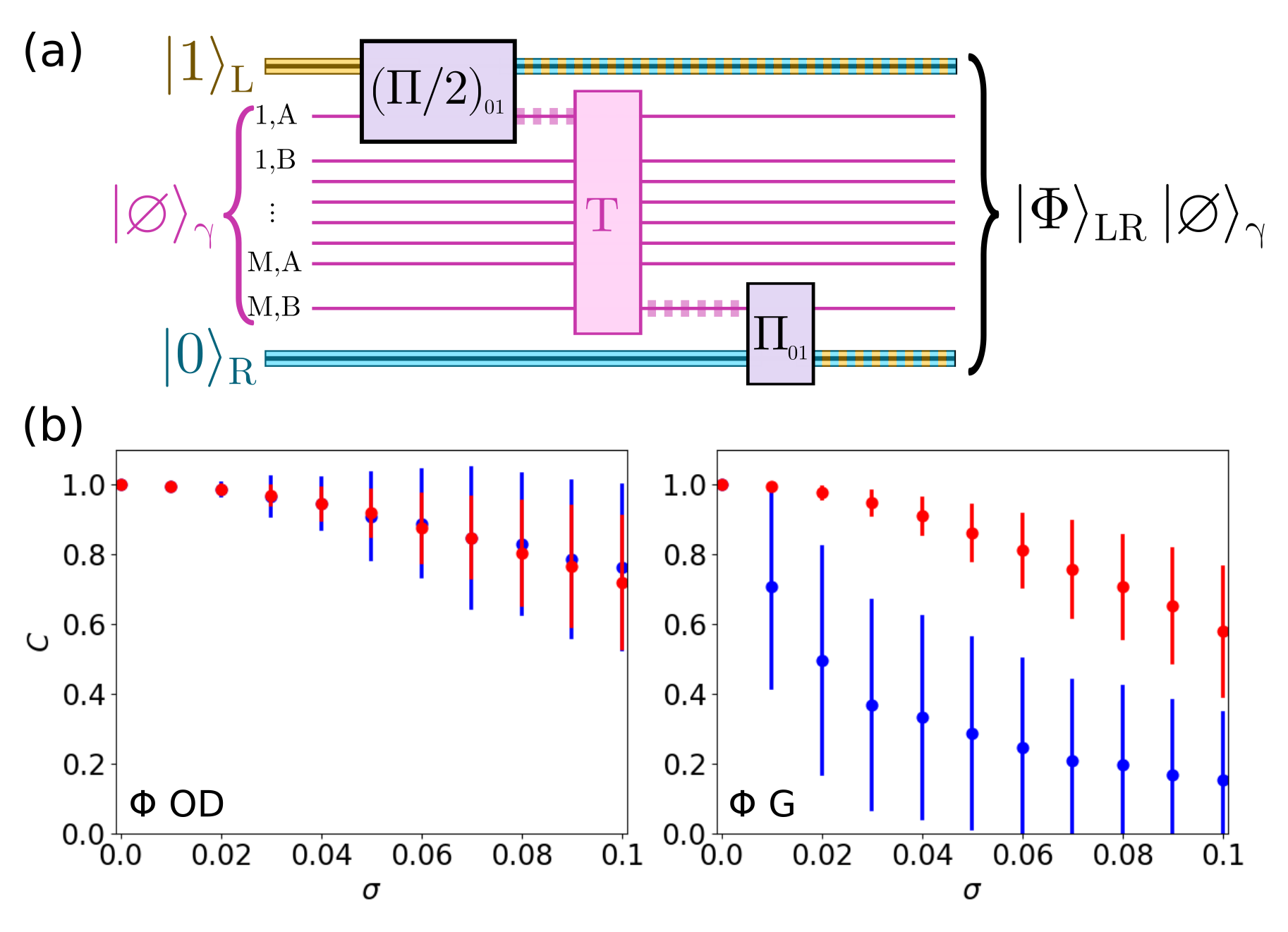}\caption{\label{fig:2Q}(a) Preparation protocols for the $\Phi$-type Bell states. Qudit states are represented as blue (0) or yellow (1). Photonic cavities are represented as thin pink lines, with an additional shading if a photon is present. Superpositions are represented with dotted lines. (b) Average concurrence and its standard deviation over 1000 realizations of the Bell states as a function of off-diagonal (OD) and general (G) disorder. We show the single-domain case ($N_{cav}=14$, blue) and the four-domain case ($N_{cav}=15$, red).}
\end{figure}

We show the results in Fig. \ref{fig:2Q} (b). In the case of off-diagonal disorder (OD), topological protection helps keep the concurrence greater than $0.7$ even at $\sigma = 0.1$ (a disorder of $10\%$ with respect to $w$). The single- and four-domain cases perform very similarly, with the single-domain case being slightly better, given that more edge states participate in the transfer in the four-domain case. However, when general disorder is considered, the four-domain case performs much better than the single-domain one. With the protecting symmetry broken, the deciding factor is now the speed of the protocol, and the presence of domain walls allows it to complete in a much shorter time. For more details on the kind and importance of the observed errors see Appendix \ref{sec:fid}.

We have considered no qubits in the domain walls for this protocol. If we included them and wanted to create entanglement between two of them, only the transfer protocol would have to be modified as detailed above.

To prepare the other kind of Bell state, $|\Psi\kt=(|00\kt+\xi|11\kt)/\sqrt{2}$, the same protocol can be used, with an additional $X$ gate acting on the second qubit at the end. We also explore another possibility, using qutrits, in Appendix \ref{sec:Psi}.

\section{Multipartite entanglement generation}\label{sec:Multi} The full connectivity of the topological photonic lattice opens the door to other protocols that, instead of entangling arbitrary pairs of qubits, entangle triplets or larger subsets of qubits. This approach displays the benefits of this nearly one-dimensional platform when it comes to scaling the number of qubits. To demonstrate this, we design protocols that prepare multipartite entangled states. Unlike in bipartite entanglement, there is no unique notion of maximally entangled multi-qubit states that, once generated with the photonic lattice, allows for the generation of any other state without the use of the photonic lattice. Instead, there are multiple families of states, each of them with a maximal state. Thus, we propose protocols for the generation of these maximal states for different families \cite{gour_wallach2013}. The simplest of these can be regarded as the W and GHZ states for three qubits \cite{dür_vidal2000}.

\subsection{W state preparation}
\begin{figure}
\includegraphics[width=1\columnwidth]{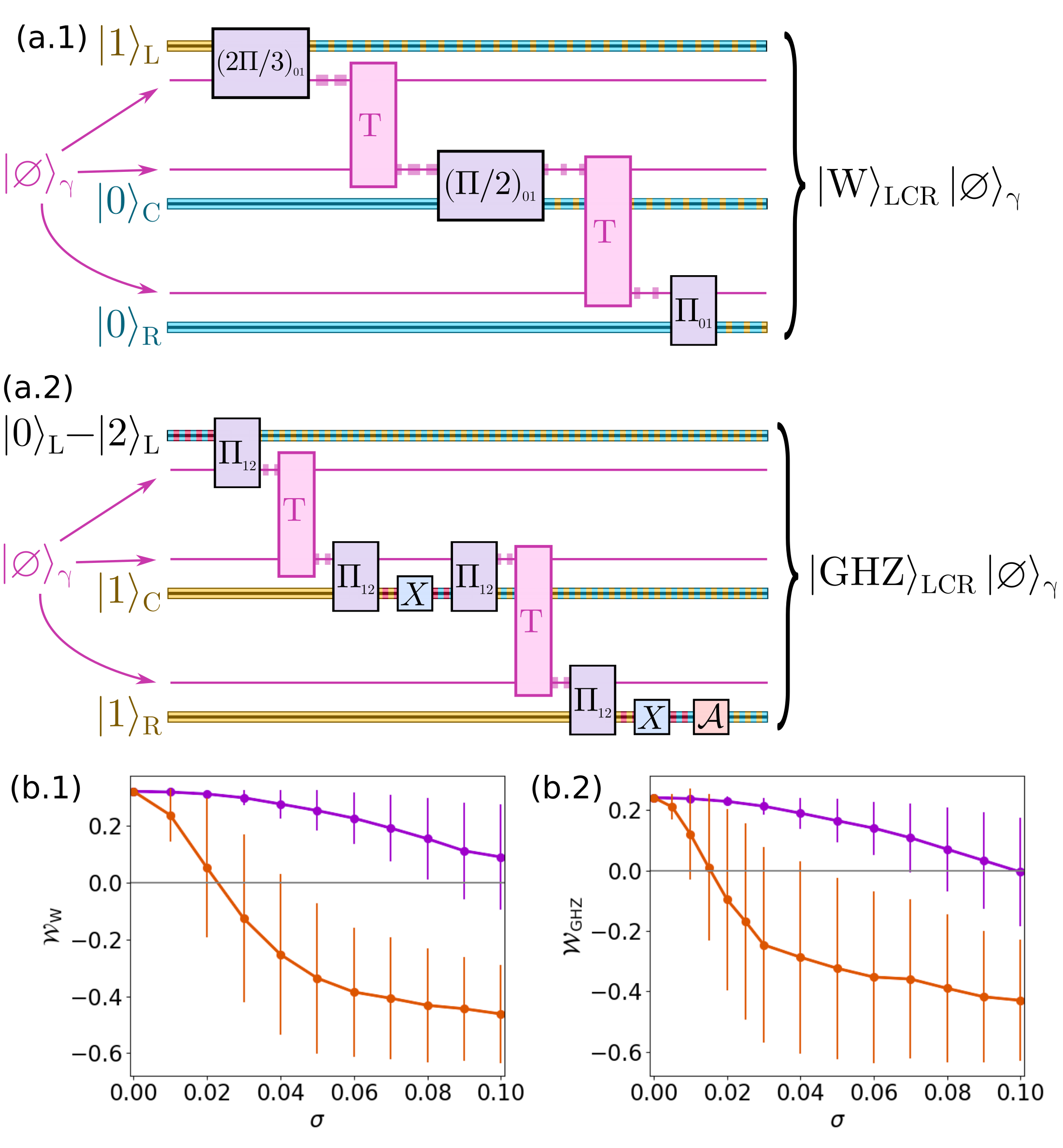}\caption{\label{fig:3Q}(a) Preparation protocols for the (a.1) W and (a.2) GHZ
states for 3 qudits: $L,C,R$. Qudit states are represented as blue
(0), yellow (1) or red (2). Photonic cavities are represented as thin
pink lines, with an additional shading if a photon is present. Superpositions
are represented with dotted lines. (b) Average values over 1000 realizations of the W witness in \refeq{w-witness} (b.1) and the GHZ witness in \refeq{ghz-witness} (b.2) as a function of off-diagonal (orange) and general (purple) disorder. The lattice has $N=2$ domains and $N_{cav}=10$ cavities.}
\end{figure}

We propose a protocol for the preparation of the W state that is similar to the previous one preparing the $\Phi$ state in \refeq{ideal-bell-state}, but with extended notation regarding the third qubit, which we label $C$ after its \textit{centered} position somewhere between the leftmost and rightmost qubits. The setup is shown in \reffig{fig:3Q} (a). With this extension in notation, the protocol is as follows.

(i) Start in state $\ket{1}_L \otimes \ket{0}_C \otimes \ket{0}_R \otimes \ket{\varnothing}_\g$. (ii) Implement a $2\pi/3$-pulse on the left qubit-cavity coupling parameter $g_L$, so that the cavity now has a probability of $2/3$ of having a photon. (iii) Perform a topological transfer to the cavity $C$ where qubit $C$ is found. (iv) Execute a $\pi/2$-pulse on qubit $C$, resulting in a qubit excitation probability of one third. Its cavity now has a $1/3$ probability of having a photon. (v) Implement a topological transfer from cavity $C$ to cavity $R$. (vi) Perform a final $\pi$-pulse on qubit $R$. Then, the final state of the qubits is:
\begin{align}
\ket{\textrm{W}}_{LCR} = \frac{1}{\sqrt{3}}\left[\ket{100} - i \ket{010}  + \ket{001}\right],
\label{ideal-w-state}
\end{align}
after taking into account all the relative phases produced by each step of the protocol. 

\subsection{GHZ state preparation}
To prepare the GHZ state, we need qutrits instead of qubits. We now suppose the gap between states $|1\kt$ and $|2\kt$, $\D\e_{12}$, can be tuned to the frequency of the cavity. State $|2\kt$ of each qutrit is used as an auxiliary state and is not present in the final result. The protocol is as follows.

(i) Start in state $(|011\kt + \eta |211\kt)/\sqrt{2}$ (with $|\eta|=1$), which has no entanglement, and an empty photonic lattice. (ii) Implement a $\pi$-pulse on the left qutrit. (iii) Perform a full transfer along the first domain. (iv) Implement a $\p$-pulse on the second qutrit. (v) Apply an $X$ gate to that qutrit\footnote{We implement operations $X$ and $\mathcal{A}$ using a sine-shaped pulse on the appropriate elements of the qudit Hamiltonian: the off-diagonal elements that flip the qutrit states and the diagonal elements that fix the ensuing phase shift. Due to the method used, a global phase of $-i$ is acquired, in addition to the implemented operation (e.g. to do an $X$ operation on the left qubit we actually implement $-i(X_L\otimes\uno_R)$).}. (vi) Repeat operations (ii-v) along the second domain: do a $\pi$-pulse on the second qutrit, realize a transfer along the second domain, and a $\pi$-pulse and $X$ gate on the third qutrit. (vii) Finish the protocol by applying operation $\mathcal{A}=|1\kt\langle2|+h.c.$ to the last qutrit [see Fig. \ref{fig:3Q} (a.2)]. The final state is:
\begin{equation}
    i|\textrm{GHZ}\kt = i(|000\kt+\eta \zeta^2 |111\kt)/\sqrt{2}.
    \label{ideal-ghz-state}
\end{equation}
In our case, $\zeta = i$, so we choose $\eta = -1$ to get no relative phase. We include the intermediate states for this and all protocols in Appendix \ref{sec:steps}.

\subsection{Performance against disorder: entanglement witnesses}
Just as with the bipartite case above, the calculations provided so far show that the system produces the maximally entangled states of the two different families of tripartite entanglement in the absence of disorder. To extend this claim to experimental settings, we remind the reader that certifying the presence of entanglement becomes more sophisticated than simple inspection of the state. Since disorder might leave some photonic probability amplitude in the domain walls, as opposed to vacuum as in the pristine case, the three-qubit reduced state is no longer pure. Then, we need to define what we mean by tripartite entanglement. In the pure state setting this is straightforward: a state is tripartitely entangled iff there is no separable subsystem; that is, $\psi \neq \phi_i \otimes \phi_{j, k}$ for any $i,j,k$, where the subindices range without repetition over the three subsystems. But in the mixed state case special care must be taken into account to rule out possibilities like
\begin{align}
\rho_{\textrm{insep}} =
P_1\rho_1 \otimes \rho_{23} 
+ P_2\rho_2 \otimes \rho_{13} 
+ P_3\rho_3 \otimes \rho_{12}
\label{non-genuine-example}
\end{align}
where the entanglement, if any, is of bipartite nature, but our information about the system is not enough to know between which qubit pair, nor which is the separable one. The subindices of each density matrix $\rho$ indicate the subsystems they span. Whenever a density matrix cannot be expressed as in \refeq{non-genuine-example}, or as incoherent mixtures of these matrices, we claim that its tripartite entanglement is genuine. For the sake of brevity, we use witnesses instead of entanglement measures to make assessments about the presence of entanglement as a function of disorder. Witnesses are functions of the state that, in our case, report the presence of genuine entanglement whenever they are positive. If they are non-positive the state might be entangled, but these quantities do not \textit{witness} it in those particular settings, hence the name. In other words, witnesses are typically built to be sensitive to a particular state and its surroundings. This is the case for the two witnesses that we consider. The first one is tailored to detect the W state in \refeq{ideal-w-state} and is
\begin{align}
    \sW_{\mathrm{W}} = - \frac{2}{3} + \textrm{tr}(\rho \Pi_{\textrm{W}} ),
    \label{w-witness}
\end{align}
where $\rho$ is the final density matrix for one realization of the protocol, and $\Pi_{\textrm{W}}$ is the projector on state $|\textrm{W}\kt_{LCR}$. Moreover, we consider another witness, tailored to the GHZ state with no relative phase in \refeq{ideal-ghz-state}:
\begin{align}
    \label{ghz-witness}
    \sW_{\textrm{GHZ}} = -\frac{3}{4} +  \textrm{tr}(\rho \Pi_{\textrm{GHZ}} ),
\end{align}
where $\Pi_{\textrm{GHZ}}$ is the projector on that state. For the derivation of these witnesses, see \cite{acin_bruss2001}. 

The average values of these witnesses as a function of disorder are presented in \reffig{fig:3Q} (b), for lattice with 10 cavities, 2 domains ($\ell = 4$) and 3 qudits. From the data we conclude that topological protection plays a fundamental role in the generation of genuine entanglement, since the witnesses remain positive for higher values of symmetry-preserving disorder (purple) than when general, symmetry-breaking disorder is present (orange). Nevertheless, the requirements placed on the system's disorder are higher with multipartite than with bipartite entanglement, since the former is a more valuable resource than the latter. In the bipartite case, some entanglement is still present when the disorder takes values as high as $0.1 w$, whereas here we need a disorder smaller than $0.05 w$ to detect multipartite entanglement. However, we must point out that, while concurrence is a measure for two-qubit entanglement, the witnesses used for three-qubit entanglement are not measures. Therefore, it could be the case that multipartite entanglement is still present for disorder higher than $0.05 w$ and further developments in the theory of genuine entanglement certification \cite{Friis2018} could detect them. We include the fidelity results for both protocols in Appendix \ref{sec:fid}, and discuss the most common errors observed that harm performance.

On a similar note, we remind the reader that this multipartite entanglement can be produced on demand between any three qubits in the scheme of \reffig{fig:model}. Additionally, the topological transfer considered here could entangle more than three qubits in a genuine way, leading to further technological applications while paving the way to become a quantum computing platform. The modifications needed to prepare generalized W and GHZ states for any number of qubits are straightforward, and we discuss them in Appendix \ref{sec:pqubits}.

\section{Probable sources of loss and decoherence}\label{sec:Loss}

Considering the experimental platform detailed in Section \ref{ssec:exp}, we now explore the main sources of loss and decoherence that might appear in the qudits and photonic lattice, and assess their effects on the protocols.

\subsection{Spontaneous qubit deexcitation}

The main source of decoherence affecting the color center qudits is spontaneous deexcitation, in which a qudit decays to the ground state without emitting a photon in the cavity. As discussed in similar setups \cite{Bello2022}, this causes the amplitude of the excited qudit states to decay in time. The fidelity of the final state then shows an exponential decay:
\begin{equation}
    f = f_0 e^{-\g t_{\textrm{tr}}},
\end{equation}
where $f_0$ is the fidelity with no decoherence, $t_{\textrm{tr}}$ is the duration of the protocol and $\g$ is the spontaneous qubit deexcitation rate.

Given that the effect depends on transfer time, the speed-up obtained in multidomain protocols can help mitigate its effects. In any case, experiments show color centers have specially long coherence times, several orders of magnitude larger than the transfer times our protocols would have in air-hole lattices \cite{Hollenbach2020,Arregui2023}. The order of magnitude\footnote{This calculation assumes a color-center coherence time of $3.8 \textrm{ ns}$ \cite{Hollenbach2020}, and a value of $w$ around $40 \textrm{ meV}$ \cite{Arregui2023}.} expected for $\gamma/w$ is around $10^{-6}$. In the absence of other noise and disorder, this would mean the fidelities for the $\Phi$ state protocol with one and four domains would be, respectively, $f_1 = 0.9997$ and $f_4 = 0.99996$. Due to its small effect, other sources of decoherence are likely to be more relevant than color center decoherence.

\subsection{Photonic scattering losses}

The sources of photonic loss in the cavity lattice can be multiple. Radiative losses, in which the photon in a cavity is lost because of radiation, are experimentally found to be negligible in air-hole cavity lattices in semiconductors \cite{Calusine2014}. Scattering losses, caused by imperfections in the air-hole lattice, are much more important. These imperfections can be perturbations on the air-hole radii, in their positions, or in the angle their walls form with the vertical.

As shown in the literature \cite{Calusine2014,Tanaka2003}, small deviations in this angle, caused by an imperfect etching process, can substantially alter the performance of the photonic lattice. Any amount of asymmetry in the vertical axis, such as the one caused by non-vertical air-holes, will couple the resonating modes of the cavities with the bulk photonic modes of the surrounding air-hole lattice. This will cause unwanted power losses into those modes, which will end up dissipating.

However, this loss will only be significant if any of the frequencies involved in the transfer is close to a bulk mode frequency of the air-hole lattice \cite{Tanaka2003}. A partial solution would be to keep the transfer frequencies---which are tuned by the control parameter $v$---lower than the bulk mode band, although sacrificing some speed in the process. In any case, the frequency of the cavities will depend strongly on the color centers used and the photonic crystal architecture, and could be more challenging to adjust.

\subsection{Off-diagonal disorder between cavities}
Finally, there is another source of decoherence, which is specific to the platform we propose. As mentioned in \cite{Hollenbach2020,Redjem2023}, controlling the exact placement and orientation of the color centers can pose a challenge, although advances are being made with controllable growth techniques \cite{Hollenbach2020,Hollenbach2022,Redjem2023}.

A possible solution would be to create the cavity lattice around the impurities after they are placed \cite{Redjem2023}. The uncertainty in their location would then induce perturbations in the position of each cavity and the distance between them, causing off-diagonal disorder. This disorder in the photonic lattice translates as a source of decoherence in the qubit subspace, as explored in Appendix \ref{sec:fid}.

However, the topological robustness against off-diagonal disorder in our protocols would greatly dampen its effects, as we show in Sections \ref{sec:Bell} and \ref{sec:Multi}. Thus, these protocols might be especially helpful in the implementation of the model.

\section{Conclusions}\label{sec:concl} In this \work, we propose a realistic
experimental setting in which fast multidomain topological protocols
can be used to quickly and reliably
generate multipartite entanglement.
We consider an easily scalable photonic lattice in silicon, with coupled qudits implemented natively with impurities, but our protocols could also be considered in other systems like superconducting or spin qubits, both of which are being investigated as platforms for photon-mediated entanglement generation \cite{Yang2016,Chan2023}.

We show how topological protection can help overcome some of the main issues associated with these setups, and we benchmark the effectiveness of that protection by studying the behaviour of entanglement measures and witnesses for increasing disorder levels. The versatility of the proposed setup and its effectiveness in entanglement distribution opens the door for quantum information applications like the generation of highly-entangled states, such as cluster states, to use as a resource for quantum computation. This constitutes an intriguing avenue for future research. Finally, the universality of operations over the protected computational states in the photonic lattice can allow an array of possible quantum operations, which will be explored further in future works.

\acknowledgements
C.E.C. was supported by the Spanish MICINN through Grant No. PID2022-139288NB-100. G.P., A.A.C. and J.Z. were supported by Spain's MINECO through Grant No. PID2020-117787GB-I00 and by CSIC Research Platform PTI-001. We also acknowledge support from National Project QTP2021-03-002. J.Z. recognizes the FPU program FPU19/03575.

\appendix

\section{Topological protection\label{sec:TopoProt}}

\begin{figure}
\includegraphics[width=1\columnwidth]{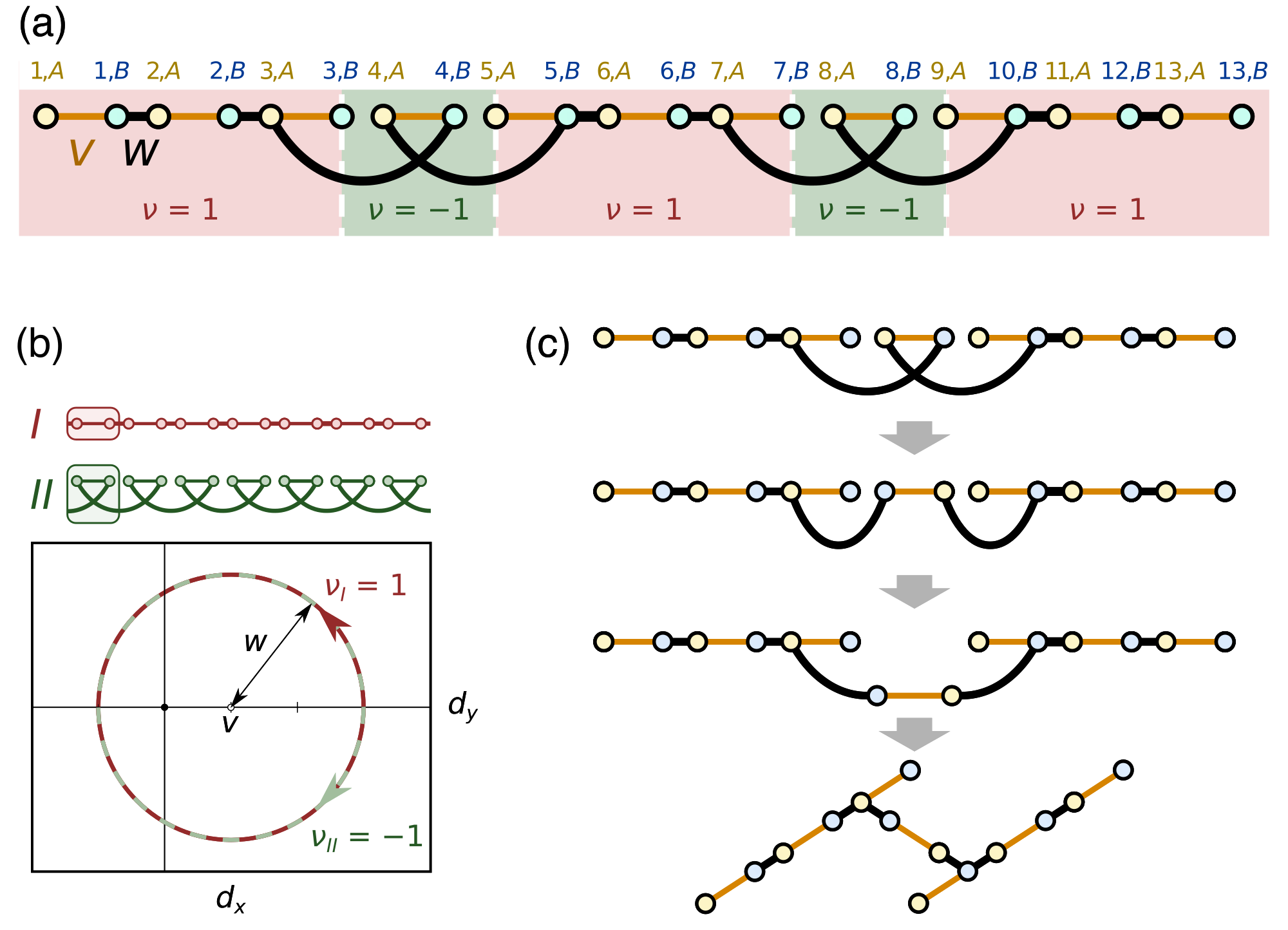}\caption{\label{fig:topo-1}(a) Possible 1D indexing of the cavities, that
allows the domains to be identified as having a winding number of
$\nu=\pm1$. (b) Path taken in the space of possible Hamiltonians
$\mathcal{H}(k)=\vec{d}\cdot\vec{\protect\s}$, where $\vec{\protect\s}$
are the Pauli matrices and $k\in BZ$, for systems $I$ and $II$
above. (c) Under deformation, the lattice in (a) is equivalent to
a three-domain lattice of the type shown in Fig. \ref{fig:model}
(a).}
\end{figure}

In order to prove the topological protection of the model analytically, we first look at the case where all $v$ terms are nonzero but smaller than $w$. As noted recently in \cite{Dias2022}, site indexing plays a vital role in the topological description of 1D models. By choosing the appropriate indexing [which we also use in Eq. (\ref{eq:Hph})], we can conceptualize the SSSH lattice as a series of topological domains with alternating winding numbers of $\nu=\pm 1$. We show this in Fig. \ref{fig:topo-1} (a) for a 5-domain model. The odd-numbered domains are regular SSH chains with a winding number of $1$, while the even domains have a winding number of $-1$. Given that each site connects to the contiguous unit cell that is further from it, instead of closer like the SSH chain, the bulk Hamiltonian $\mathcal{H}(k)$ is identical to the SSH chain but with $k\to -k$. That reverses the path traced by $\mathcal{H}(k)$ in the space of possible Hamiltonians, changing the sign of its winding number, as shown in Fig. \ref{fig:topo-1} (b). The model we use here is completely equivalent to the one in the main text, as shown in Fig. \ref{fig:topo-1} (c). The latter does not use long-range hopping terms.

\begin{figure}
\includegraphics[width=1\columnwidth]{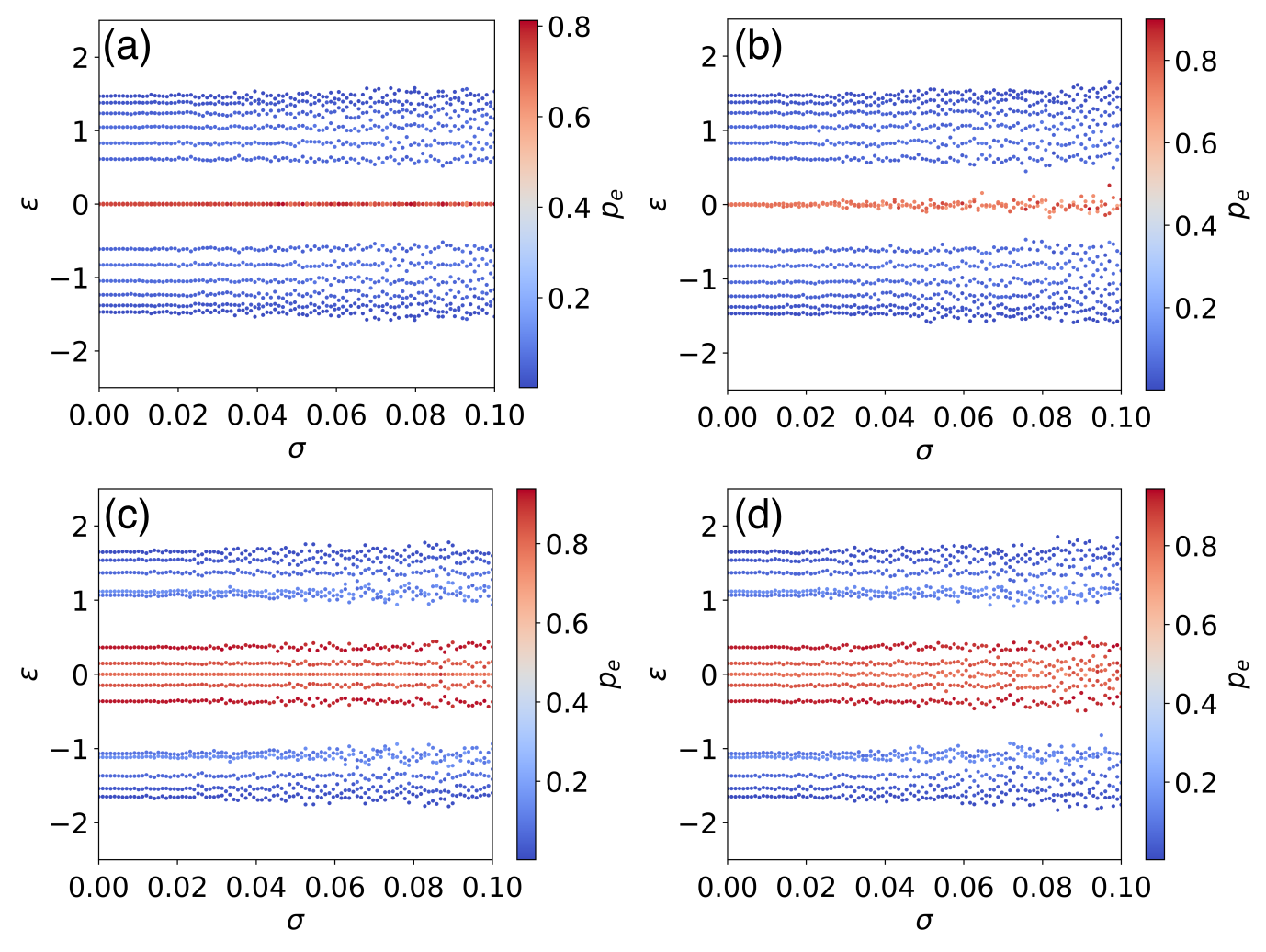}\caption{\label{fig:spectra}Spectrum of the single- (a,b) and four-domain (c,d) models over a single realization, for increasing values of both off-diagonal (a,c) and general disorder (b,d). The color of each point represents an approximate probability of finding it in a topological state (see main text), $p_e$. Topological edge states are shown in red, and trivial states in blue.}
\end{figure}

The difference in topological invariant between the two sides of a boundary dictates the number and chirality of protected states. The ends of the system will have one protected state ($\sL$ and $\sR$), and each wall will have two ($\mathcal{S}$ and $\mathcal{P}$), see Fig. \ref{fig:model} (b). The chiralities (i.e. the sublattice that supports the states) alternate between successive walls. These states will be protected as long as the chiral symmetry is preserved, and so they can be used to store information ($\sL, \sR$ and $\mathcal{S}$) or help in the transfer ($\mathcal{P}$). This is the general case.

Another possible case, that we use to prepare the 3-qubit states, is that one of the domains is made trivial due to a high value of $v=v_{\mathrm{bar}}\gg w$. In that case, the interface between that domain and its topological neighbour will only have one protected state ($\mathcal{S}$), see Fig. \ref{fig:model} (c). The transfer process will not continue into the trivial domain, which acts like a sort of barrier.

During a multidomain transfer, the intermediate qubit cavities are left unconnected, as shown in Fig. \ref{fig:model} (d). This case can be seen as a multidomain SSH chain, the topology of which has been discussed in the literature at length \cite{Su1980, Munoz2018, Qi2021, Zurita2023}. The topological protection is also guaranteed under this point of view.

Having proven the presence of topological protection using the bulk-boundary correspondence, we now stop to closely examine the meaning of that protection in each of our protocols. To that end, we study numerically the behaviour of the spectrum of the photonic lattice with respect to the level and type of disorder in the system.

In each realization, the presence of disorder will change the eigenvalues of the system randomly, and this effect will increase with the disorder strength $\sigma$. If a semi-infinite SSH chain was considered, the topological protection would imply that the only end state, which would be its own chiral partner, would be pinned at zero as long as the chiral symmetry is preserved.

In a long, single-domain SSH model, such as the 14-cavity dimerized chain used in this work, the two end modes hybridize and acquire a small positive and negative energy. The two energy eigenstates are now each other's chiral partner, and so they have opposite energies in the presence of chiral symmetry. This situation can be seen in Fig. \ref{fig:spectra} (a).

\begin{figure}
\includegraphics[width=1\columnwidth]{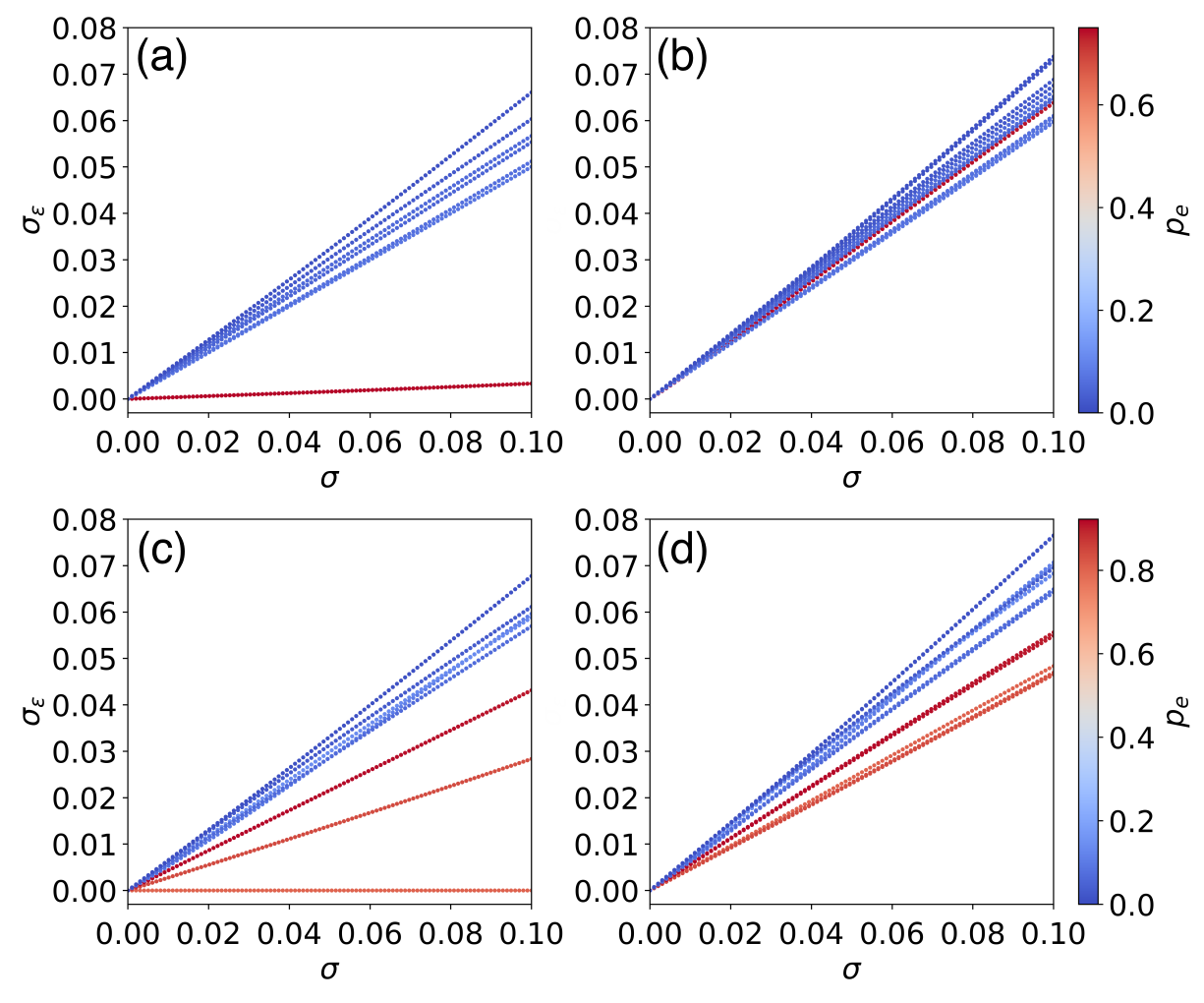}\caption{\label{fig:stds}Standard deviation of each state in the spectrum of the single- (a,b) and four-domain (c,d) models over $10^4$ realizations, for increasing values of both off-diagonal (a,c) and general disorder (b,d). The color of each point represents an approximate probability of finding it in a topological state (see main text), $p_e$. Topological edge states are shown in red, and trivial states in blue.Topological protection in the presence of undisturbed chiral symmetry is more robust in the single-domain case, but it is also more fragile when the symmetry is broken.}
\end{figure}

This small energy, however, is the cause of the exponentially long durations achieved with single-domain chains \cite{Zurita2023}. To fight this, we increase the number of domains, making them closer to each other. As a consequence, their hybridization energies are dramatically larger, as shown in Fig. \ref{fig:spectra} (c) for the four-domain, 15-cavity SSH model corresponding to the connected sites in Fig. \ref{fig:model} (d). They are still clearly separated from the bulk, but their energy fluctuations are larger than in the single-domain case.

To study the amplitude of these fluctuations, we calculate the standard deviation of the energy of each photonic lattice eigenstate, $\sigma_\epsilon$. We group the states in different realizations by their order in energy, and we can differentiate between topological and bulk states by calculating the probability $p_e$ of finding the particle at the topological end modes\footnote{We use the completely dimerized form of the topological modes for simplicity to find $p_e$, but this is enough to clearly differentiate between bulk and edge states.}.

We show the results in Fig. \ref{fig:stds} for the single-domain, 14-site model (a,b) and the four-domain, 15-site one (c,d), for both kinds of disorder and $10^4$ realizations. Each of the realizations has a fixed disorder profile, scaled to match $\sigma$. As expected, all edge states show a smaller $\sigma_\epsilon$ than the bulk states for off-diagonal disorder, although the single-domain case fluctuates much less, as expected. In the general disorder case, however, the single-domain model end modes immediately lose all protection, while the four-domain model states continue being the more constant ones. This shows that, for a fixed length, the topological protection of a multidomain model is weaker than the single-domain case but, in the more realistic case of general disorder, can resist for larger disorder levels.

The other hallmark of topological protection, bulk-boundary decoupling, is studied in depth in Appendix \ref{sec:fid}, where we break down the results of all disorder simulations.

\begin{figure*}[t]
\includegraphics[width=\textwidth]{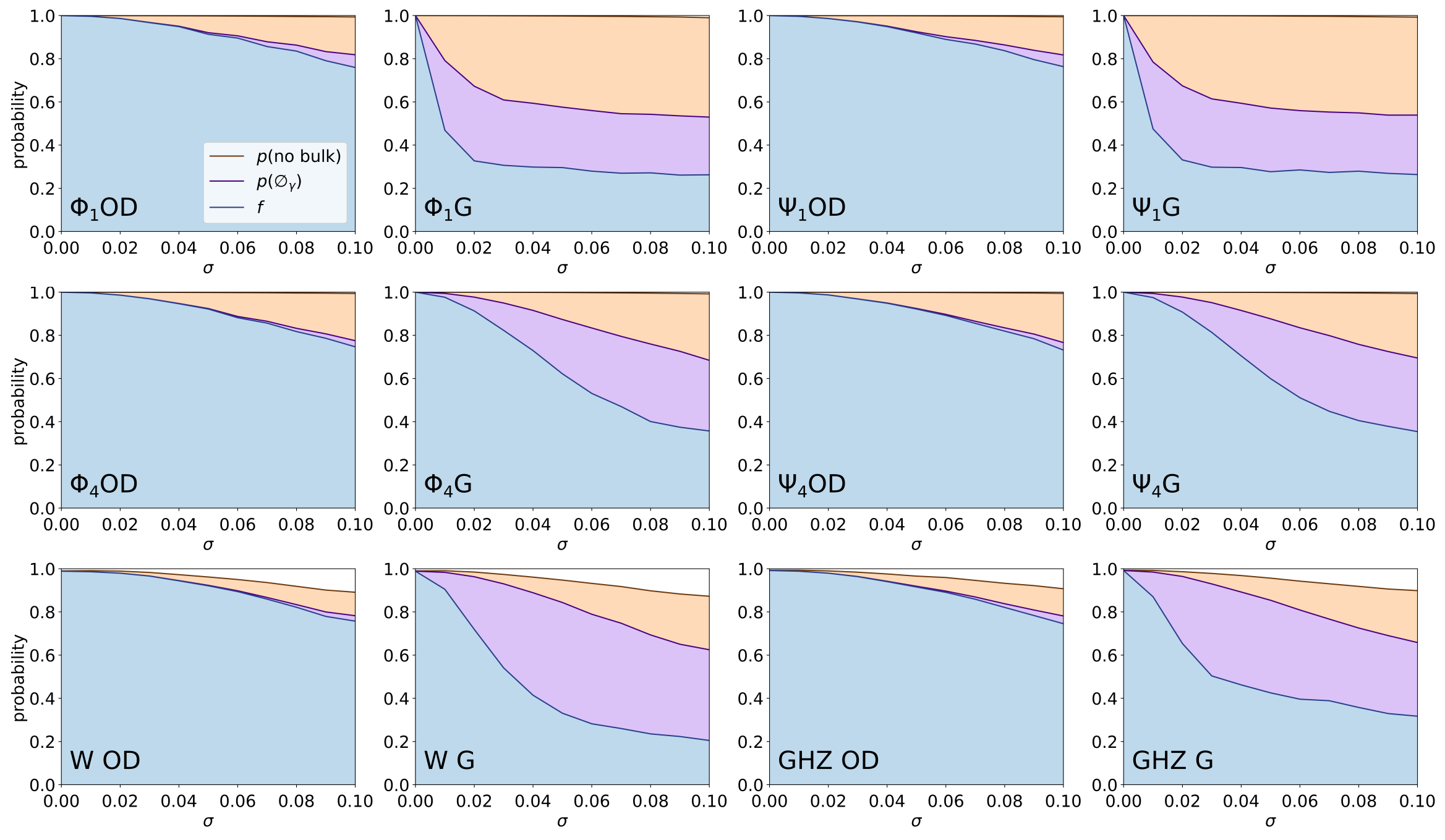}\caption{\label{fig:sand}Disaggregated probabilities for all disordered protocols, i.e. the single- and four-domain $\Phi$ and $\Psi$ protocols, and the W and GHZ ones; both with off-diagonal (OD) and general (G) disorder. The fidelity $f$, the probability of finding no photon in the cavities $p(\varnothing_\gamma)$, and that of finding no photon in the bulk states $p(\textrm{no bulk})$ are represented in blue, purple and brown lines, respectively. The purple shaded region shows the probability of a failed final state with no photon, caused by imperfect qudit-cavity coupling operations. The orange shaded region shows the probability of a failed final state with a photon in the topological states, while the white region above $p(\textrm{no bulk})$ is the probability of leakage into the bulk. The latter is almost zero, except in the protocols that use trivial domains as barriers (W and GHZ).}
\end{figure*}

\section{Details on the control pulses}\label{sec:pulses}
The pulses used have three parts: an initial preparation time where the control parameter is switched on slowly during an interval of $t_{\mathrm{prep}}$, a time where the parameter is kept constant, and a final interval lasting $t_{\mathrm{prep}}$ where it is switched off slowly. All pulses used are symmetric, and $t_{\mathrm{prep}}$ is chosen so that $t_{\mathrm{prep}} \gg w^{-1}$, in order for the process to be adiabatic with respect to the energy gap between bulk and edge states. The same kind of pulses, although in different models, were used in \cite{Zurita2023}. We keep $w=1$ throughout.
 
If we set $t=0$ at the start of the pulse, the formula for the pulses we use is:
\begin{equation}
    v(t)  =  \begin{cases}
        v_0 \sin^2 (\Omega t) &\textrm{if } 0 \le t \le   t_{\mathrm{prep}} \\
        v_0 &\textrm{if }   t_{\mathrm{prep}} < 0 \le t_{\mathrm{tr}} - t_{\mathrm{prep}} \\
        v_0 \sin^2 [\Omega (t - t_{\mathrm{tr}})] &\textrm{if } t_{\mathrm{tr}} - t_{\mathrm{prep}} < t \le t_{\mathrm{tr}},
    \end{cases} \nonumber
\end{equation}
where the maximum $v_0$ can either be $v_0=v_{\mathrm{tr}}<w$, for the topological domains that allow the transfer, or $v_0=v_{\mathrm{bar}}\gg w$, for the trivial domains that stop the transfer (we will rename $t_{\mathrm{prep}}$ to $t_{\mathrm{bar}}$ in that case). Both kinds of pulses are nested, i.e. the barrier is lifted before ramping up $v_{\mathrm{tr}}$, and it is lowered after switching $v_{\mathrm{tr}}$ off. 

The links leading to the cavities with impurities, marked as $u_k$ in Fig. \ref{fig:model} (a), will either be switched off or tuned at the same time as a neighboring domain, depending on whether that qudit is the start or end of the transfer. This can be seen in detail for several transfer protocols in Fig. \ref{fig:W}.

When the number of domains is $n_d>2$, the control parameters need to be different in different domains for the transfer time to be minimal. The height of the pulses are chosen symmetrically with respect to the center of the transfer length, and they are found numerically once the $v_{\mathrm{tr}}$ (the height at the first and last domains) is fixed. This is explained in depth in \cite{Zurita2023}.

For the qubit-cavity couplings $g_k(t)$ that create the $(n\pi/m)$-pulses, we use the protocol above, but setting $t_{\mathrm{prep}}$ equal to half the pulse duration, so there is no constant part. We fix $g_0=0.5$ in all cases, and then find the pulse duration using the formula:
\begin{equation}
    t_{\mathrm{coup}} = \frac{n\pi}{mg_0}.
\end{equation}
We include in Table \ref{tab:params} the parameters for all protocols used in this work, when applicable: domain number $n_d$, domain length $\ell$, sites involved in each transfer $L$, control parameter heights $v_{\mathrm{tr}}$ and $v_{\mathrm{tr}}^\prime$ (for the central domains in the 4-domain case), barrier height $v_{\mathrm{bar}}$, and the times for preparation ($t_{\mathrm{prep}}$), barrier preparation ($t_{\mathrm{bar}}$) and total duration ($t_{\mathrm{tr}}$).

\begin{table}
    \centering \scriptsize
    \begin{tabular}{c|c|c|c|c|c|c|c|c|c} \hline  
           States \rule{-5.5pt}{0pt}&\rule{-5.5pt}{0pt} $n_d$ \rule{-5.5pt}{0pt}&\rule{-5.5pt}{0pt}$\ell$\rule{-5.5pt}{0pt}&\rule{-5.5pt}{0pt}$L$\rule{-5.5pt}{0pt}&\rule{-5.5pt}{0pt}$v_{\mathrm{tr}}$\rule{-5.5pt}{0pt}&\rule{-5.5pt}{0pt}  $v_{\mathrm{tr}}^{\prime}$\rule{-5.5pt}{0pt}&\rule{-5.5pt}{0pt}  $v_{\mathrm{bar}}$\rule{-5.5pt}{0pt}&\rule{-5.5pt}{0pt} $t_{\mathrm{prep}}$ \rule{-5.5pt}{0pt}&\rule{-5.5pt}{0pt} $t_{\mathrm{bar}}$ \rule{-5.5pt}{0pt}&\rule{-5.5pt}{0pt} $t_{\mathrm{tr}}$\rule{-5.5pt}{0pt} \\ \hline \rule{-5.5pt}{0pt}   
           Bell 1 dom.  \rule{-5.5pt}{0pt}&\rule{-5.5pt}{0pt} 1 \rule{-5.5pt}{0pt}&\rule{-5.5pt}{0pt} 12 \rule{-5.5pt}{0pt}&\rule{-5.5pt}{0pt} 16  \rule{-5.5pt}{0pt}&\rule{-5.5pt}{0pt} 0.5 \rule{-5.5pt}{0pt}&\rule{-5.5pt}{0pt} -- \rule{-5.5pt}{0pt}&\rule{-5.5pt}{0pt} --  \rule{-5.5pt}{0pt}&\rule{-5.5pt}{0pt} 25 \rule{-5.5pt}{0pt}&\rule{-5.5pt}{0pt} 15 \rule{-5.5pt}{0pt}&\rule{-5.5pt}{0pt} 304.0\rule{-5.5pt}{0pt} \\ \hline \rule{-5.5pt}{0pt}   
           Bell 4 doms. \rule{-5.5pt}{0pt}&\rule{-5.5pt}{0pt} 4 \rule{-5.5pt}{0pt}&\rule{-5.5pt}{0pt} 4  \rule{-5.5pt}{0pt}&\rule{-5.5pt}{0pt} 15 \rule{-5.5pt}{0pt}&\rule{-5.5pt}{0pt} 0.5 \rule{-5.5pt}{0pt}&\rule{-5.5pt}{0pt} 0.38 \rule{-5.5pt}{0pt}&\rule{-5.5pt}{0pt} --  \rule{-5.5pt}{0pt}&\rule{-5.5pt}{0pt} 20 \rule{-5.5pt}{0pt}&\rule{-5.5pt}{0pt} 15 \rule{-5.5pt}{0pt}&\rule{-5.5pt}{0pt} 45.3\rule{-5.5pt}{0pt} \\ \hline \rule{-5.5pt}{0pt}   
           W, GHZ       \rule{-5.5pt}{0pt}&\rule{-5.5pt}{0pt} 1 \rule{-5.5pt}{0pt}&\rule{-5.5pt}{0pt} 4  \rule{-5.5pt}{0pt}&\rule{-5.5pt}{0pt} 6 \rule{-5.5pt}{0pt}&\rule{-5.5pt}{0pt} 0.5 \rule{-5.5pt}{0pt}&\rule{-5.5pt}{0pt} -- \rule{-5.5pt}{0pt}&\rule{-5.5pt}{0pt} 30 \rule{-5.5pt}{0pt}&\rule{-5.5pt}{0pt} 7 \rule{-5.5pt}{0pt}&\rule{-5.5pt}{0pt} 15 \rule{-5.5pt}{0pt}&\rule{-5.5pt}{0pt} 25.2\rule{-5.5pt}{0pt} \\
    \end{tabular}
    \caption{Parameters for all transfer protocols used, and the states prepared with it.}
    \label{tab:params}
\end{table}

\section{Acquired phases and phase-shift protocols}\label{sec:phases}

The analytical form of the topological modes involved in a transfer is:

\begin{align}
    |\sL\kt &= -\mathcal{N}_\sL \sum_{j=1}^{\ell/2} \left(-\frac{w}{v}  \right)^{-j}|j,A\kt\\
   |\mathcal{P}_k\kt &= \mathcal{N}_\mathcal{P} \left[\sum_{j=j_0-\ell/2}^{j_0 - 1} \left(-\frac{w}{v_l}  \right)^{j-j_0+1} |j,\alpha(k) \kt,   \right.\nonumber\\
    &\left. - \sum_{j=j_0 + 1}^{j_0 + \ell/2}   \left(-\frac{w}{v_r}  \right)^{j_0-j+1} |j,\alpha(k)\kt,    \right] \\
    |\sR\kt &= -\mathcal{N}_\sR \sum_{j=N_c - \ell/2}^{N_c} \left(-\frac{w}{v}  \right)^{-j+N_c+1}|j,\alpha(N)\kt,
\end{align}
where $j_0 = \ell k /2 + 1$ is the unit cell where domain wall $k$ is located, $\alpha(k)=A$ ($B$) for even (odd) $k$, $N$ is the number of domains and $N_c$ is the total number of unit cells. In the $\mathcal{P}$ states, we allow the control parameter on the left ($v_l$) and right ($v_r$) of the wall to be different, as they will be in general.

If the initial or final states of the transfer are located in a domain wall instead of the left or right ends (thus being $\mathcal{S}$ states instead of $\sL$ or $\sR$ states), their form is identical except for a translation to take them to said wall.

Using these and the Hamiltonian in Eq. (\ref{eq:Hph}), we obtain that, after a transfer across $n_d$ domains with $\ell$ sites each, the final state will have a phase factor of:
\begin{equation}
    \zeta(n_{d},\ell)=-[(-1)^{\ell/2+1}i]^{n_{d}}, \label{eq:zeta}
\end{equation}
when compared to the original state.

As mentioned in the main text, shifts in the relative phases of the maximally entangled states do not affect the amount of entanglement. However, they could be changed natively if necessary by temporarily adding an on-site potential in the relevant cavity after all transfers have been made. We use the same pulses as for the control parameters $v$, fixing the height of the pulse $\epsilon_0$ and the preparation time $t_{\mathrm{prep}}^{(\varphi)}$ and then finding the total duration as:
\begin{align}    
    t_{T}^{(\varphi)} = t_{\mathrm{prep}}^{(\varphi)} - \frac{|\varphi|}{\epsilon_0},
\end{align}
where $\varphi$ is the desired phase shift. The sign of the pulse is determined by the sign of $\varphi\in[-\pi,\pi)$. All this ensures the area under the pulse provides the intended shift.

\begin{figure}
\includegraphics[width=1\columnwidth]{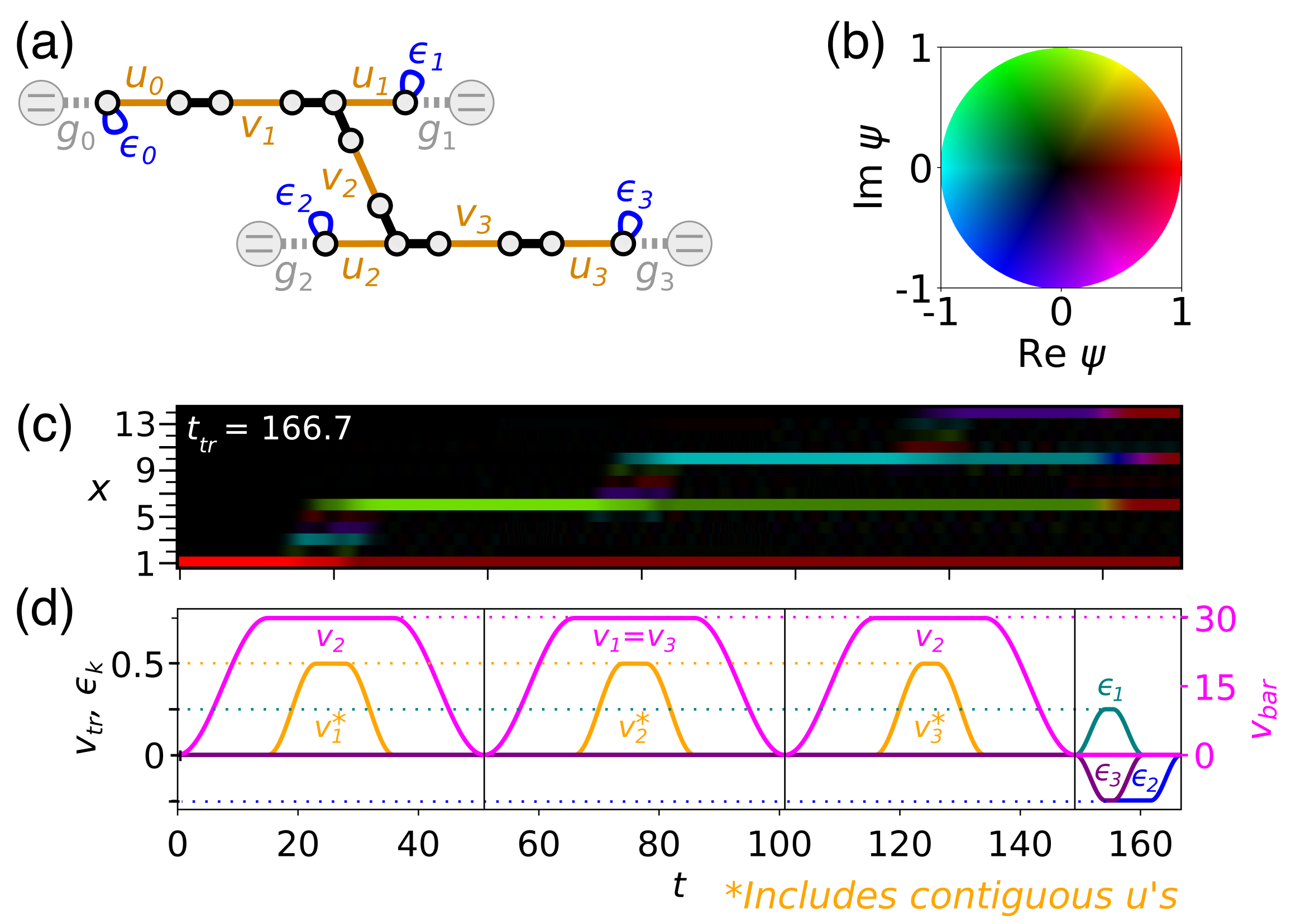}\caption{\label{fig:W}Partial transfer processes and phase shifts. (a) The photonic
lattice considered, a three-domain SSSH chain. (b) The legend used
in (c), with color labeling the magnitude and phase of the photon
wavefunction at each cavity. (c) Transfer processes needed to obtain
$|\psi\protect\kt_{\protect\g}=(|1,A\protect\kt+|3,B\protect\kt+|5,A\protect\kt+|7,B\protect\kt)/2$,
chosen as an example. (d) The value of the control parameters during
the protocol, see (a). Each color refers to different parameters in
each part, separated with dotted lines. Orange lines ($v_{\mathrm{tr}}$) also
include the contiguous $u_{k}$'s, in order for the transfer to be
induced properly; pink lines ($v_{\mathrm{bar}}$) do not. The phase shifts
are induced using on-site energies $\epsilon_{k}$.
%This protocol can be used to prepare the corresponding W-type state, $|W\protect\kt_{Q}=\frac{1}{2}(|1000\kt + |0100\kt + |0010\kt + |0001\kt)$, if the lattice is coupled to external qubits.
}
\end{figure}

To show this, we distribute a photon equally between the four boundary cavities of a 3-domain SSSH model, showing the acquired phases. Then, the relative phases are dynamically shifted using on-site energies, until we eliminate all relative phases (see Fig. \ref{fig:W}). The final state is then $(|1,A\kt+|3,B\kt+|5,A\kt+|7,B\kt)/2$. This phase-shift operation can be used in any entanglement generation protocol to choose the relative phases.

\section{Intermediate states for each protocol} \label{sec:steps}
For completeness, we include in this section the intermediate states of the system during each of the protocols.

The states at each step in the $\Phi$ state preparation protocol are:

{\small
\begin{flalign*}
    &|\f_0\kt = |10\vac\kt_{LR\g}&&\\
    &\down (\Pi/2)^{(L)}_{01}&&\\
    &|\f_1\kt = \frac{1}{\sqrt{2}}[|10\vac\kt-i|00\sL\kt]&&\\
    &\down T_{\sL\sR}&&\\
    &|\f_2\kt = \frac{1}{\sqrt{2}}[|10\vac\kt-i\z|00\sR\kt]&&\\
    &\down (\P)^{(R)}_{01}&&\\
    &|\f_3\kt = \frac{1}{\sqrt{2}}[|10\vac\kt-\z|01\vac\kt]=|\Phi\kt_{LR}\otimes|\vac\kt_\g,&&   
\end{flalign*}
}
where $\zeta$ is the phase acquired in the transfer.

To prepare a $\Psi$ state, an $X_R$ gate must be performed on the final state of the protocol above, producing:
{\small
\begin{flalign*}
    |\psi\kt=-\frac{i}{\sqrt{2}}[|11\vac\kt-\z|00\vac\kt] = |\Psi\kt_{LR}\otimes|\vac\kt_\g,&&
\end{flalign*}
}
where the global phase shift is a consequence of our implementation for $X_R$, as mentioned in the main text.

For the W state protocol, the intermediate states are:
{\small
\begin{flalign*}
    &|w_0\kt = |100\vac\kt_{LCR\g}&&\\
    &\down (2\Pi/3)^{(L)}_{01}&&\\
    &|w_1\kt = \frac{1}{\sqrt{3}}[|100\vac\kt-i\sqrt{2}|000\sL\kt]&&\\
    &\down T_{\sL\sS}&&\\
    &|w_2\kt = \frac{1}{\sqrt{3}}[|100\vac\kt-i\z\sqrt{2}|000\sS\kt]&&\\
    &\down (\P/2)^{(C)}_{01}&&\\
    &|w_3\kt = \frac{1}{\sqrt{3}}[|100\vac\kt-\z|010\vac\kt-i\z|000\sS\kt]&&\\
    &\down T_{\sS\sR}&&\\
    &|w_4\kt = \frac{1}{\sqrt{3}}[|100\vac\kt-\z|010\vac\kt-i\z^2|000\sR\kt]&&\\
    &\down \Pi^{(R)}_{01}&&\\
    &|w_5\kt = \frac{1}{\sqrt{3}}[|100\vac\kt-\z|010\vac\kt-\z^2|001\vac\kt]=&&\\
    &=|W\kt_{LCR}\otimes|\vac\kt_\g.&&
\end{flalign*}
}
where $\zeta$ is the phase acquired in each transfer.

And finally, for the GHZ state protocol:
{\small
\begin{flalign*}
    &|g_0\kt = \frac{1}{\sqrt{2}}[|011\vac\kt_{LCR\g}+\y|211\vac\kt_{LCR\g}]&&\\
    &\down \Pi^{(L)}_{12}&&\\
    &|g_1\kt = \frac{1}{\sqrt{2}}[|011\vac\kt-i\y|111\sL\kt]&&\\
    &\down T_{\sL\sS}&&\\
    &|g_2\kt = \frac{1}{\sqrt{2}}[|011\vac\kt-i\y\z|111\sS\kt]&&\\
    &\down \P^{(C)}_{12}&&\\
    &|g_3\kt = \frac{1}{\sqrt{2}}[|011\vac\kt-\y\z|121\vac\kt]&&\\
    &\down X_C&&\\
    &|g_4\kt = \frac{-i}{\sqrt{2}}[|001\vac\kt-\y\z|121\vac\kt]&&\\
    &\down \Pi^{(C)}_{12}&&\\
    &|g_5\kt = \frac{-i}{\sqrt{2}}[|001\vac\kt+i\y\z|111\sS\kt]&&\\
    &\down T_{\sS\sR}&&\\
    &|g_6\kt = \frac{-i}{\sqrt{2}}[|001\vac\kt+i\y\z^2|111\sR\kt]&&\\
    &\down \P^{(R)}_{12}&&\\
    &|g_7\kt = \frac{-i}{\sqrt{2}}[|001\vac\kt+\y\z^2|112\vac\kt]&&\\
    &\down X_R&&\\    
    &|g_8\kt = \frac{-1}{\sqrt{2}}[|000\vac\kt+\y\z^2|112\vac\kt]&&\\
    &\down \sA_R&&\\
    &|g_9\kt = \frac{i}{\sqrt{2}}[|000\vac\kt+\y\z^2|111\vac\kt]=&&\\
    &=i|GHZ\kt_{LCR}\otimes|\vac\kt_\g.&&
\end{flalign*}
}

\section{Fidelity and bulk-edge decoupling}\label{sec:fid}

In this section, we study the average fidelity of the protocols. For a given realization, it is obtained as:
\begin{equation}
    f = \langle \psi_{0} | \rho_Q |\psi_{0} \rangle,
\end{equation}
where $ |\psi_{0} \rangle$ is the intended final state, and $\rho_Q$ is the density matrix of the qudit system at the end of the realization, after tracing out the photonic degrees of freedom. For every protocol considered in this work, we obtain a fidelity higher than 0.99 for the pristine system, which is the current error-correction threshold \cite{Fowler2012}. We include the $\Phi$, W and GHZ states prepared as in the main text, and the $\Psi$ state, prepared as detailed in Appendix \ref{sec:Psi}.

In order to get the full picture, we also study two more metrics: the probability of the photonic lattice ending up empty, $p(\varnothing_\gamma)$, and the probability of finding the bulk subspace completely empty, $p(\textrm{no bulk})$. Of course, $p(\textrm{no bulk}) > p(\varnothing_\gamma) > f$, because the intended final states have no photons left in the lattice, and if the lattice is empty, the bulk sites are too. If we plot the average of these quantities over all 1000 realizations as a function of disorder for each case, an interesting picture appears.

The average fidelity over 1000 realizations, shown in blue in Fig. \ref{fig:sand}, is smaller in the general disorder case due to the absence of topological protection, but also due to the fact that on-site fluctuations can affect Rabi flopping protocols, like the ones we use, more than off-diagonal disorder can. Additionally, when comparing Fig. \ref{fig:2Q}(b) and Fig. \ref{fig:sand}, it can be observed that when the fidelity drops down in the general disorder case for the single-domain protocols, it retains a higher value than the concurrence does. This is due to the fact that some failed protocols that do not create entanglement at all can still return a nonzero fidelity. For example, a final qudit state of $|00\rangle$ will still result in a fidelity of 0.5 when trying to implement state $|\Psi\rangle$, even though no entanglement is created in the process.

We now consider the average probability of finding the photonic lattice empty but not ending up at the intended state, which can be calculated as $p(\varnothing_\gamma) - f$, and is represented by the height of the purple sector in Fig. \ref{fig:sand}. Given the protocols and the kind of disorder we consider (i.e. only in the photonic lattice), this result can only be caused by errors in the qudit-photon-coupling operations. It turns out these, which are a form of Rabi flopping themselves, are much more affected by on-site fluctuations than off-diagonal disorder, and so this sector increases greatly when considering general disorder with respect to off-diagonal disorder. An interesting research direction to follow would be to look for ways to protect these operations against on-site disorder, using topology or other methods.

Finally, the orange sector of the figures represent the average probability of finding a photon in the lattice at the end of the protocol, but only with support on the topological state manifold. It has been obtained as:
\begin{equation}
    p(\textrm{edge}) = (\sum_k \bra k \rho_{ph} \ket k)^2,
\end{equation}
where $\rho_{ph}$ is the density matrix of the photonic degrees of freedom. The probability $p(\textrm{no bulk})$ is then found as $p(\textrm{no bulk}) = p(\varnothing_\gamma) + p(\textrm{edge})$. We can see that it is extremely close to 1, even in the presence of general disorder. This shows that the bulk-edge decoupling caused by the topology of the system is remarkably robust. This has been reported for other topological systems in the literature \cite{Lang2017}. The three-qudit cases show a slightly smaller decoupling due to the use of trivial domains as barriers, which can allow some photon amplitude to leak into the bulk. This problem could be mitigated by decoupling the domains altogether by switching off some $w$ amplitudes, but we decided to keep them constant throughout in this work for simplicity.

Finally, we want to acknowledge the existence of trivial boundary states in the domain walls (in the fully dimerized case, for example, they are located in the 3-site strong-strong defects, and spanned by the second site in them and the 0-phase superposition of the first and third sites in them). These states could, in principle, be a hindrance for the robustness of the protocols, but their effect was shown to be negligible in all two-qubit protocols, and extremely small in the three-qubit ones (with a probability of less than 0.02 of finding a photon in those states with a general disorder of 0.1).

\section{GHZ-like protocol for Bell states}\label{sec:Psi}

\begin{figure}[ht]
\includegraphics[width=1\columnwidth]{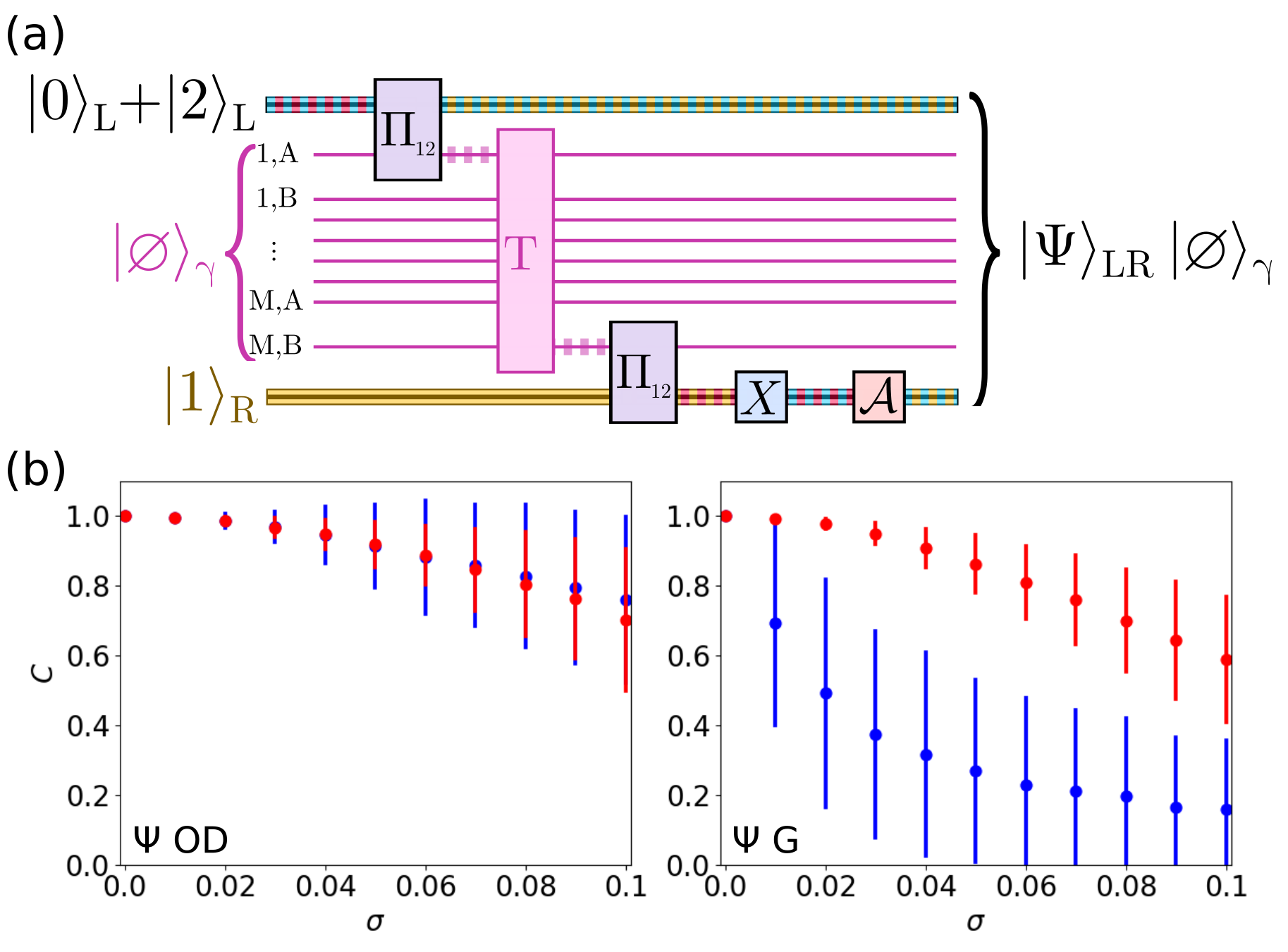}\caption{\label{fig:Psi}(a) GHZ-like protocol, adapted to create state $|\Psi\kt$. States 0,1 and 2 are shown in blue, yellow and red, respectively. A photon in a cavity is shown as purple shading. (b) Concurrence as a function of disorder strength for off-diagonal disorder (left) and general disorder (right) for an SSH chain with one domain and 14 sites (blue), and for a four-domain chain with 15 sites (red).}
\end{figure}

In this section, we include for completeness the concurrence results of the preparation of the $|\Psi\kt = |00\kt+\zeta|11\kt$ state using the protocol described in the main text for the GHZ state, restricted to two qutrits. We show the protocol and results in Fig. \ref{fig:Psi}. Given that it only uses a single transfer, just like the protocol used in the main text for $|\Phi\kt$, its results are remarkably similar.

\section{Generalization to any number of qubits} \label{sec:pqubits}
In this section, we generalize the W and GHZ protocols to prepare a $p$-qubit generalized W or GHZ state. The pulses on the generalized W state protocol depend on the number of qubits. The steps to prepare it are: (i) Start in state $|1\cdots 0\rangle$, with an empty photonic lattice. (ii) Implement a $(p-1)\pi/p$-pulse on the left qubit-cavity coupling parameter. (iii) Realize a full transfer to the next qubit cavity. (v) Repeat steps (ii-iii) for each qubit successively, now implementing a $(q-1)\pi/q$-pulse in step (ii), with $q=p-1,\ldots,2$. (iv) After the final transfer, implement a $\pi$-pulse on the last qubit, achieving a uniform probability distribution among all intervening states. The final state is:

\begin{align}
    &|\textrm{W}_p\kt = \frac{1}{\sqrt{p}}\left( |100\cdots 0\kt -\zeta_{0,1}|010\cdots 0\kt  \right.  \\
    &\left. - \zeta_{0,1}\zeta_{1,2}|001\cdots 0\kt + \cdots -\zeta_{0,1}\cdots \zeta_{p-1, p} |000\cdots 1\kt    \right) ,  \nonumber
\end{align}
where $\zeta_{k,k^\prime}$ is the phase acquired in the transfer from the cavity housing qubit $k$ to the cavity housing qubit $k^\prime$, which can be obtained using Eq. (\ref{eq:zeta}).

To prepare the generalized GHZ state we need qutrits instead of qubits, just like in the three-qubit version, and we suppose that the gap between qutrit states $|1\kt$ and $|2\kt$ can be tuned to the frequency of the cavity. The steps required are: (i) Start in state $(|01\cdots 1\kt + \eta |21 \cdots 1\kt)/\sqrt{2}$, with $|\eta|=1$, which has no entanglement, and an empty photonic lattice. (ii) Implement a $\pi$-pulse on the left qutrit. (iii) Realize a full transfer along the first domain. (iv) Implement a $\p$-pulse on the next qutrit. (v) Apply an $X$ gate to that qutrit. (vi) Repeat operations (ii-v) for each additional qutrit $k$, that is, $\Pi_{k-1}T_{k-1,k}\Pi_{k}X_{k}$. (vii) Finish the protocol by applying the operation $\mathcal{A}=|1\kt\langle 2|+h.c.$ to the last qutrit. The final state is:

\begin{align}
    |\textrm{GHZ}_p\kt = \frac{1}{\sqrt{2}}\left( |00\cdots 0\kt + (-i)^{2(p-1)}\eta|11\cdots 1\kt     \right).
\end{align}

The value of $\eta$ in the initial state can be chosen to fix the relative phase. Both protocols can be easily modified via the qudit-cavity coupling operations to change the relative weights between their final components, and the relative phases in the W state can be adjusted with phase gates or using the phase-shift protocols in Appendix \ref{sec:phases}. We include all intermediate states in Appendix \ref{sec:steps}.

%\bibliographystyle{quantum}
%\bibliography{TopoBell.bib}

\end{document}